\documentclass[12pt,a4paper,roman]{aa}  
\usepackage{rotating} 
\usepackage[english]{babel}
\usepackage{setspace}
\usepackage{siunitx}
\usepackage[utf8]{inputenc}
\usepackage{polski}
\usepackage[T1]{fontenc}
\usepackage{graphicx}
\usepackage{wrapfig}
\usepackage{makecell}
\usepackage{booktabs}
\usepackage{txfonts}
\usepackage{amsmath}
\usepackage{float}
\usepackage{natbib}
\usepackage{graphicx}
\usepackage{enumitem}
\usepackage{dblfloatfix}
\usepackage{txfonts}
\usepackage{subcaption}
\usepackage{multirow}
\usepackage{caption}

\usepackage{amsmath}
\usepackage{float}
 \usepackage[table,xcdraw]{xcolor}
\usepackage[hidelinks,colorlinks=true,linkcolor=blue,citecolor=blue]{hyperref}
\DeclareUnicodeCharacter{2212}{-}
\usepackage{xcolor}

\def \magperarcsec{mag arcsec$^{-2}$}
\def \mue{$\bar{\mu}_{eff}$}
\def \re{$r_{1/2}$}
\def \msun{$\,M_{\odot}\,$}

\begin{document}

   \title{Shedding Light on Low Surface Brightness Galaxies in Dark Energy Survey with Transformers}

   \author{H. Thuruthipilly \inst{1}\and
          Junais\inst{1}\and
          A. Pollo\inst{1,2}\and
          U. Sureshkumar\inst{2,3}\and
          M. Grespan \inst{1}   \and
          P. Sawant \inst{1}   \and
          K. Małek \inst{1}\and
          A. Zadrozny\inst{1}}

   \institute{National Centre for Nuclear Research, Warsaw, Poland\\
             \email{Hareesh.Thuruthipilly@ncbj.gov.pl},\\
             \email{Junais@ncbj.gov.pl}, 
             \email{Agnieszka.Pollo@ncbj.gov.pl}
             \and
            Jagiellonian University, Krak\'{o}w, Poland 
            \and Wits Centre for Astrophysics, School of Physics, University of the Witwatersrand, Johannesburg, South Africa.}

   \date{Received XXX / Accepted YYY}
  \abstract
   {Low surface brightness galaxies (LSBGs) which are defined as galaxies that are fainter than the night sky, play a crucial role in understanding galaxy evolution and cosmological models. Upcoming large-scale surveys like Rubin Observatory Legacy Survey of Space and Time (LSST) and Euclid are expected to observe billions of astronomical objects. In this context, using semi-automatic methods to identify LSBGs would be a highly challenging and time-consuming process and demand automated or machine learning-based methods to overcome this challenge.}
   { We study the use of transformer models in separating LSBGs from artefacts in the data from the Dark Energy Survey (DES) data release 1. 
   Using the transformer models, we then search for new LSBGs from the DES that the previous searches may have missed. 
   Properties of the newly found LSBGs are investigated, along with an analysis of the properties of the total LSBG sample in DES.  }
   {We created eight different transformer models and used an ensemble of these eight models to identify LSBGs. This was followed by a single-component S\'ersic model fit and a final visual inspection to filter out false positives.}
   {Transformer models achieved an accuracy $\sim94\%$ in separating the LSBGs from artefacts. In addition, we identified 4\,083 new LSBGs 
   in DES, adding an additional $\sim17\% $ to the LSBGs already known in DES. This also increased the number density of LSBGs in DES to 5.5 deg$^{-2}$. The new LSBG sample consists of mainly blue and compact galaxies. We performed a clustering analysis of the LSBGs in DES using an angular two-point auto-correlation function and found that LSBGs cluster more strongly than their high surface brightness counterparts. This effect is driven by the red LSBG. We associated 1310 LSBGs with galaxy clusters and identified 317 among them as ultra-diffuse galaxies (UDGs). We found that these cluster LSBGs  
   are getting bluer and larger in size towards the edge of the clusters when compared with those in the centre.}
   {Transformer models have the potential to be on par with convolutional neural networks as state-of-the-art algorithms in analysing astronomical data. The significant number of LSBGs identified from the same dataset using a different algorithm highlights the substantial impact of the methodology on finding LSBGs. The reported number density of LSBGs is only a lower estimate and can be expected to increase with the advent of surveys with better image quality and more advanced methodologies.}


   \maketitle
%
\section{Introduction}\label{introduction}

Low-surface-brightness galaxies (LSBGs) are most often defined as galaxies with a central surface brightness fainter than the night sky or galaxies with \textit{B}-band central surface brightness $\mu_0(B)$ below a certain threshold value. In literature, the threshold values of $\mu_0(B)$ for classifying a galaxy as LSBG vary among different works, ranging from $\mu_0(B)\geq23.0$ mag arcsec$^{-2}$ \citep{Bothun} to $\mu_0(B) \geq 22.0$ mag arcsec$^{-2}$ \citep{Burkholder}.

It is estimated that the LSBGs only contribute a few percentages ($<10\%$) to the local luminosity and stellar mass density of the observable universe \citep{Bernstein, Driver, Hayward, Martin}. However, LSBGs are considered to account for a significant fraction $(30\%\sim60\%)$ of the total number density of galaxies \citep{McGaugh, Bothun, Neil, Haberzettl, Martin}, and as much as $15\%$ of the dynamical mass content of the universe \citep{Driver, Minchin}.
These numbers imply that LSBGs can contribute significantly to our understanding of the physics of galaxy evolution and cosmological models. However, as their name indicates, LSBGs are very faint systems, and due to the observational challenges in detecting them, LSBGs remain mostly an unexplored realm. 

In recent years, despite the observational challenges, advances in digital imaging have improved our ability to detect LSBGs. The first known and the largest LSBG to be 
identified and verified is Malin 1, serendipitously discovered by \citet{Malin} during a survey of galaxies of low surface brightness in the Virgo cluster. Notably, Malin 1 is the largest spiral galaxy known until today \citep[e.g.,][]{Impey, Junais, Galaz1}. 
Current searches for LSBGs have shown that they exhibit a wide range of physical sizes \citep{Greene} and can be found in various types of environments, ranging from satellites of 
local nearby galaxies \citep{Danieli, Cohen}, ultra-faint satellites of the Milky Way \citep{McConnachie, Simon}, galaxies found in the field \citep{Leisman, Prole}, to  members of massive galaxy clusters like Virgo \citep{Mihos1, Mihos2, Junais2022} and Coma \citep{Dokkum, Koda}.

LSBGs also consist of several sub-classes based on their physical size, surface brightness and gas content. Ultra-diffuse galaxies (UDGs) represent a subclass of LSBGs characterized by their considerable size, comparable to that of Milky Way-like galaxies, yet exhibiting very faint luminosities akin to dwarf galaxies. Although the term `UDG' was coined by \citet{Dokkum}, such galaxies were identified in several earlier studies in the literature \citep{Sandage,McGaugh_bothun, Dalcanton, Conselice_udgs}.
Similarly, giant LSBGs (GLSBGs) form another sub-class of LSBGs that are extremely gas-rich (M$_{\rm HI} > 10^{10}$ \msun{}), faint and extended \citep{Sprayberry1995, Saburova2023}. The formation and evolution of extreme classes like UDGs and GLSBGs are still debated \citep{Amorisco, Cintio, Saburova2021, Benavides, Laudato}.

To comprehend the formation of various types LSBGs in different environments,
studying them extensively across different environments (galaxy clusters vs field) over a large area of the sky is crucial. Recently, \citet{Greco} detected 781 LSBGs in the Hyper Suprime-Cam Subaru Strategic Program (HSC SSP) in a blind search covering around 200 deg$^2$ of the sky from the Wide layer of the HSC SSP. Similarly, in a recent study, \citet{Tanoglidis1} utilised a support vector machine (SVM) and visual inspection to analyse the first three years of data from the Dark Energy Survey (DES). They identified more than 20\,000 LSBGs, which is currently the largest LSBG catalogue available.

A common feature observed in both of these untargeted searches for LSBGs was the significant presence of low-surface brightness artefacts. As pointed out in \citet{Tanoglidis1}, these artefacts predominantly consist of diffuse light from nearby bright objects, galactic cirrus, star-forming tails of spiral arms and tidal streams.
These artefacts typically pass the simple selection cuts based on photometric measurements and often make up the majority of the LSBG candidate sample. These contaminants need to be removed, which is often accomplished using semi-automated methods with a low success rate and visual inspection, which is more precise but time-consuming.

For example, in HSC SSP, \citet{Greco} applied selection cuts on the photometric measurements from {\tt SourceExtractor} \citep{sxtractor}. This led to the selection of 20\,838 LSBG candidates. Using a galaxy modelling pipeline based on {\tt imfit} \citep{Erwin}, the sample size was subsequently reduced to 1\,521. However, after the visual inspection, only 781 candidates were considered confident LSBGs, which is around 4\% of the preliminary candidate sample and 50\% of the sample selected by the pipeline. Similarly, in DES, \citet{Tanoglidis1} shortlisted 419\,895 LSBG candidates using the selection cuts on {\tt SourceExtractor} photometric measurements. After applying a feature-based machine learning (ML) classification (SVM) on the photometric measurements, the candidate sample was further reduced to 44\,979 objects. However, a significant number of false positives still remained, and only 23\,790 were later classified as confident LSBGs. Therefore, these numbers indicate that the occurrence of  LSBGs in these methods is roughly 5\% for the initial selection and 50\% for the subsequent selection. 

The upcoming large-scale surveys, such as Legacy Survey of Space and Time \citep[LSST;][]{Ivezi__2019} and Euclid \citep{scaramella2021euclid}, are expected to observe billions of astronomical objects. In this scenario, relying solely on photometric selection cuts or semi-automated methods such as galaxy model fitting would not be practical to identify LSBGs confidently. Furthermore, the accuracy of the classification methodology between LSBGs and artefacts must be improved to achieve meaningful results. Hence, this situation demands more effective and efficient automation methodologies for the searches of LSBGs.

Recently, the advancements in deep learning have opened up a plethora of opportunities and have been widely applied in astronomy. Particularly for analysing astronomical images, convolutional neural networks (CNNs) have emerged as a state-of-the-art technique. For example, the CNNs have been used for galaxy classification \citep{2019PASP..131j8002P}, galaxy merger identification  \citep{Pearson}, supernova classification \citep{2017ApJ...836...97C} and finding strong gravitational lenses  \citep{ Schaefer_2018,2019MNRAS.487.5263D, rojas2021strong}. One of the fascinating features of CNNs is their ability to directly process the image as input and learn the image features, making them one of the most popular and robust architectures in use today. Generally, the learning capacity of a neural network increases with the number of layers in the network. The first layers of the network learn the low-level features, and the last layers learn more complex features \citep{russakovsky2015imagenet,simonyan2015deep}. 

One of the main requirements for creating a trained CNN is a sufficiently large training dataset that can generalise the features of the data we are trying to analyse. Recently, \citet{Tanoglidis2} utilised a catalogue of over 20\,000 LSBGs from DES to classify LSBGs from artefacts using a CNN for the first time and achieved an accuracy of 92\% and a true positive rate of 94\%.

While CNNs have been the dominant choice for analysing image data in astronomy, the current state-of-the-art models for computer vision are transformers. Transformers were initially introduced in natural language processing (NLP) as an attention-based model \citep{vaswani2017attention}. The fundamental concept behind the transformer architecture is the attention mechanism, which has also found a broad range of applications in machine learning \citep{zhang2019selfattention,fu2019dual,ramachandran2019standalone,zhao2020exploring,tan2021explicitly}.  In the case of NLP, attention calculates the correlation of different positions of a single sequence to calculate a representation of the sequence. Later the idea was adapted to computer vision and has been used to produce state-of-the-art models for various image processing tasks like image classification \citep{Wortsman}, and image segmentation \citep{Chen}.

Generally, two categories of transformers are present in the literature. The first type integrates both CNNs and attention to perform the analysis. An example of this type is the Detection Transformer (DETR) proposed for end-to-end object detection by \citet{carion2020endtoend}. The key idea behind using CNNs and Transformers together is to leverage the strengths of both architectures. CNNs excel at local feature extraction, capturing low-level details, and spatial hierarchies, while attention layers excel at modelling global context and long-range dependencies. The second class of transformers is the models that do not use a CNN and operate entirely based on self-attention mechanisms. An example of this type is the Vision transformer (ViT) proposed for object classification by \citet{dosovitskiy2020image}. ViTs have demonstrated remarkable performance in image classification tasks and have surpassed the accuracy of CNN-based models on various benchmark datasets \citep{dosovitskiy2020image, Jiahui, Wortsman}.

Even though transformers have been introduced very recently in astronomy, they have already found a wide variety of applications. For example, the transformer models have been used to detect and analyse strong gravitational lensing systems \citep{Hareesh, Hareesh2, Huang_tranformer, jia}, representing light curves which can be used further for classification or regression \citep{Allam}, and classifying multi-band light curves of different supernovae (SN) types \citep{Pimentel}.

In this paper, we explore the possibilities of transformers in classifying LSBGs from artefacts in DES and compare the performance of transformers with the CNNs presented in \cite{Tanoglidis2}. We also use the transformer models to look for new LSBGs that the previous searches may have missed. For comparison purposes, throughout this work, we follow the LSBG definition from \citet{Tanoglidis1}, based on the $g$-band mean surface brightness ($\bar{\mu}_{eff}$) and the half-light radii ($r_{1/2}$). We consider LSBGs as galaxies with $\bar{\mu}_{eff} > 24.2$ \magperarcsec{} and $r_{1/2} > 2.5\arcsec$.

The paper is organised as follows. Sect. \ref{data} discusses the data we used to train our models and look for new LSBGs. Sect. \ref{method} provides a brief overview of the methodology used in our study, including the models' architecture, information on how the models were trained, and the details about the visual inspection. The results of our analysis are presented in Sect. \ref{results}. A detailed discussion of our results and the properties of the newly identified LSBGs are analysed in Sect. \ref{discs} and Sect. \ref{newlsbs} respectively. Further analysis of the clustering of LSBGs is presented in Sect. \ref{clustering} and a detailed discussion on the UDGs, which are identified as a subsample of LSBGs, is presented in Sect. \ref{UDGS}. Sect. \ref{conlclsuin} concludes our analysis by highlighting the significance of LSBGs, the impact of methodology in finding LSBGs, and the future prospect with LSST.

\section{Data}\label{data}

\subsection{Dark Energy Survey}

The Dark Energy Survey (DES; \citealt{DR1, DESDR2}) is a six-year observing program (2013-2019) covering $\sim5000 \text{ deg}^2$ of the southern Galactic cap in the optical and near-infrared regime using the Dark Energy Camera (DECam) on the 4-m Blanco Telescope at the Cerro Tololo Inter-American Observatory (CTIO). The DECam focal plane comprises 62 2k $\times$ 4k charge-coupled devices (CCDs) dedicated to science imaging and 12 2k $\times$ 2k CCDs for guiding, focus, and alignment. 
The DECam field-of-view covers $3\text{ deg}^2$ with a central pixel scale of $0.263$ arcsec pixel$^{-1}$ \citep{Flaugher}. To address the gaps between CCDs, DES utilises a dithered exposure pattern \citep{Neilsen} and combines the resulting individual exposures to form coadded images, which have dimensions of 0.73 $\times$ 0.73 degrees \citep{Morganson}. The DES has observed the sky in \textit{grizY} photometric bands with approximately 10 overlapping dithered exposures in each filter (90 sec in \textit{griz}-bands and 45 sec in \textit{Y}-band).

\subsection{DES DR1 and the gold catalogue}\label{preselection}
In this work, we use the image data from the dark energy survey data release 1 (DES DR1; \citealt{DR1}) and the DES Y3 gold coadd object catalogue (DES Y3\_gold\_2\_2.1) obtained from the first three years of the DES observations \citep{Sevilla}. The DES DR1 comprises optical and near-infrared imaging captured over 345 different nights between August 2013 and February 2016. The median $3\sigma$ surface brightness limits of the $g$, $r$, and $i$-bands of DES DR1 are 28.26, 27.86, and 27.37 mag arcsec$^{-2}$, respectively \citep{Tanoglidis1}. It is worth mentioning that the DES source detection pipeline has not been optimised for detecting large, low surface-brightness objects \citep{Morganson}. Thus, the above-mentioned surface brightness values can be considered as the limits for detecting faint objects in each band. The gold catalogue shares the same single image processing, image coaddition, and object detection as the DES DR1. The objects in the gold catalogue were detected using {\tt SourceExtractor} \citep{sxtractor} and have undergone selection cuts on minimal image depth and quality, additional calibration, and deblending. The median coadd magnitude limit of the DES Y3 gold object catalogue at a signal-to-noise ratio (S/N) = 10 is $g = 24.3$ mag, $r = 24.0$ mag, \text{ and } $i = 23.3$ mag \citep{Sevilla}. The gold catalogue contains around 319 million astronomical objects, which we used for searching LSBGs in DES. For a detailed review and discussion on the data from the DES, please refer to \citet{DR1} and \citet{Sevilla}.

We reduced the number of objects processed in our study using preselections similar to \citet{Greco} and \citet{Tanoglidis1}. We first removed objects classified as point-like objects in the gold catalogue,  based on the \textit{i}-band {\tt SourceExtractor SPREAD\_MODEL} parameter and  {\tt EXTENDED\_CLASS\_COADD} as described in \citet{Tanoglidis1}. 
In addition, we constrained the \textit{g}-band half-light radius ({\tt FLUX\_RADIUS\_G}) and surface brightness ({\tt MUE\_MEAN\_MODEL\_G}) within the range of $2.5{\arcsec}<r_{1/2}<20{\arcsec}$ and $24.2<\bar{\mu}_{eff}<28.8$ mag arcsec$^{-2}$, respectively.
Furthermore, we also limited our sample to objects with colors (using the  {\tt MAG\_AUTO} magnitudes) in the range: 
\begin{gather}
     -0.1 < g - i < 1.4,\\
     (g - r) > 0.7 \times (g - i) - 0.4,\\
     (g - r) < 0.7 \times (g - i) + 0.4.
 \end{gather}
These color cuts are based on \citet{Greco} and \citet{Tanoglidis1}. As mentioned by \citet{Greco}, these color requirements will remove the spurious detections due to optical artefacts detected in all bands and blends of high-redshift galaxies.
Finally, we also restricted the axis ratio ({\tt B\_IMAGE/A\_IMAGE}) of each object to be greater than 0.3 to remove the artefacts like the highly elliptical diffraction spikes. Our complete selection criteria were based on the selection criteria presented in Appendix B of \citet{Tanoglidis1}. After the preliminary selections using the {\tt SourceExtractor} parameters from the DES Y3 gold catalogue, our sample contains 419\,784 objects. 

\subsection{Training data}\label{traing_data}
All of the trained, validated, and tested models in this study used the labelled dataset of LSBGs and artefacts identified from DES by \citet{Tanoglidis1}. Below, we briefly summarise the primary steps taken by \citet{Tanoglidis1} in constructing the LSBG catalogue.
\begin{enumerate}[label=(\roman*)]
    \item The {\tt SourceExtractor} parameters from the DES Y3 gold coadd object catalogue presented by \citet{Sevilla} were used to create the initial selection cuts, as discussed in Sect. \ref{preselection}. 
    \item The candidate sample was further reduced using an SVM to classify artefacts and LSBGs. The SVM was trained with a manually labelled set of $\sim 8\,000$ objects (640 LSBGs) and using the {\tt SourceExtractor} parameters as features for learning.
    \item From the candidate sample generated through SVM, over 20\,000 artefacts were excluded upon visual inspection. Most of the rejected objects that had passed SVM's feature-based selection were found to be astronomical artefacts (such as galactic cirrus, star-forming extensions of spiral arms, and tidal streams) rather than instrumental artefacts (such as scattered light emitted by nearby bright objects) during visual inspection.
    \item Objects that passed the visual inspection were subjected to S\'ersic model fitting and Galactic extinction correction. Following this, new selection cuts were applied to the updated parameters, and the final LSBG catalogue containing 23\,790 LSBGs was created.
\end{enumerate}

\begin{figure*}[h]
  \centering
  \begin{subfigure}{0.475\linewidth}
\centering
\includegraphics[width=\linewidth,keepaspectratio]{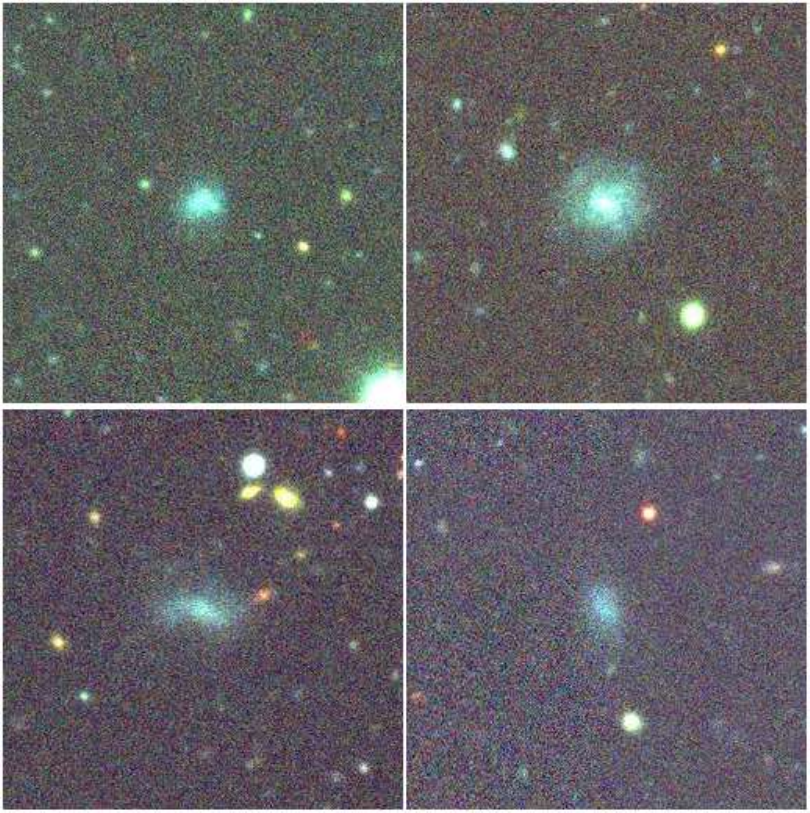}
  \caption{}
\label{lsb_train*}
\end{subfigure}
\begin{subfigure}{0.475\linewidth}
\centering
\includegraphics[width=\linewidth,keepaspectratio]{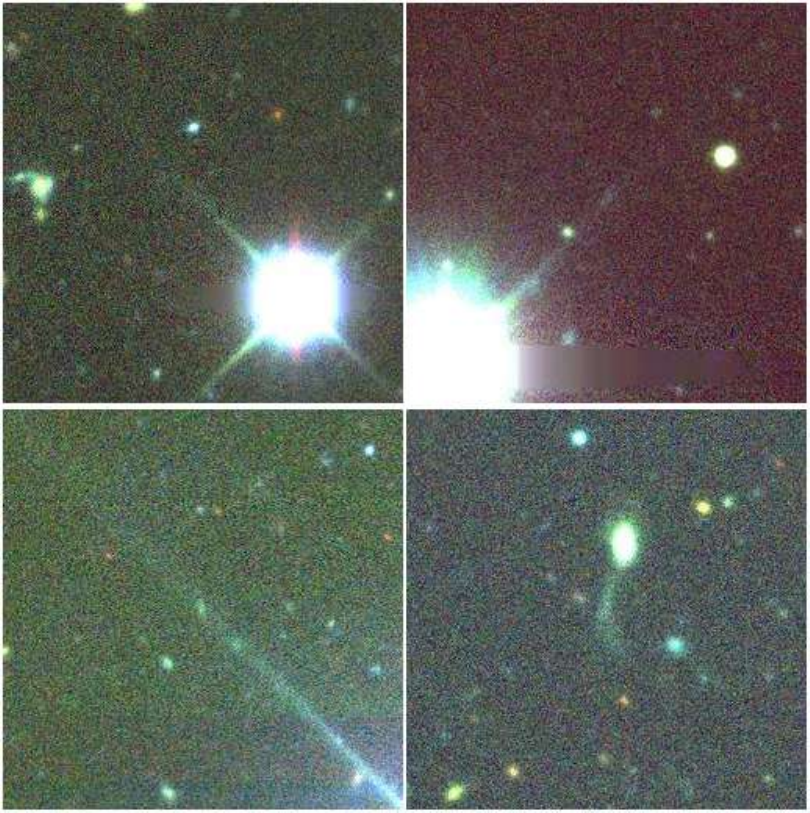}
\caption{}
\label{art_train}
\end{subfigure}
\caption{Four examples of LSBGs (\ref{lsb_train*}) and artefacts (\ref{art_train}) used in the training data. Each image of the LSBG and artefact corresponds to a $ 67.32 {\arcsec} \times  67.32 {\arcsec}$ region of the sky. Images were generated by combining the $g,r$ and $z$ bands using {\tt APLpy} package \citep{Robitaille}.}
\label{fig:training}
\end{figure*}

 For training our classification models, we selected LSBGs from the LSBG catalogue as the positive class (label - 1) and the objects rejected in the third step (visual inspection) by \citet{Tanoglidis1} as the negative class  (label - 0). The catalogues for the positive and negative classes are publicly available, and we used these catalogues to create our training dataset\footnote{\url{https://github.com/dtanoglidis/DeepShadows/blob/main/Datasets}}. The selection of the artefacts and LSBGs for training was random, and after selection, we had 18\,474 artefacts and 23\,103 LSBGs.  
However, when we further inspected these LSBGs and artefacts, we found that there were 797 objects belonging to both classes. After conducting a thorough visual examination, we identified that these are, in fact, LSBGs that had been mistakenly categorized as artefacts in the publicly accessible artefact catalogue. However, we avoided these 797 objects from our training set to avoid contamination and ambiguity among classes during training. We generated multi-band cutouts for each object in the flexible image transport system (FITS) format using the cutout service provided in the DES public data archive.
Each cutout corresponds to a $ 67.32 {\arcsec} \times  67.32 {\arcsec}$ ($256 \times 256 $ pixels) region of the sky and is centred at the coordinates of the object (LSBG or artefact). We resized the cutouts from their initial size to $64 \times 64$ pixels to reduce the computational cost. The cutouts of $g,r \text{ and } z$-bands were stacked together to create the dataset for training the models. Examples of LSBGs and artefacts used for training the model are shown in Fig.  \ref{fig:training}.
Our training catalogue contains 39\,983 objects, out of which 22\,306 are LSBGs and 17\,677 are artefacts. Before training, we randomly split the full sample into a training set, a validation set and a test set, each consisting of 35\,000, 2\,500, and 2\,483 objects, respectively. 

\section{Methodology}\label{method}
\subsection{Transformers and Attention}
As mentioned in Sect. \ref{introduction}, the central idea behind every transformer architecture is attention. Before applying attention, the input sequence is transformed into three vectors in multi-head attention: query ($Q)$, key ($K$), and value ($V$). The dot product between the query and key vectors is used to obtain attention scores. The attention scores are then used to weight the value vector, producing a context vector that is a weighted sum of the value vectors. For our work, the vectors ($Q$, $V$ and $K$) are identical, and this method is termed self-attention. This approach enables the transformer to model long-range dependencies and capture complex patterns in the input sequence. Mathematically, the attention function is defined  as 
\begin{equation}\label{attention}
 \centering
 \text{Attention}(Q,K,V) = \text{softmax} \left( \frac{QK^T}{\sqrt{d_k}} \right) V,
\end{equation}
where $Q, K, V$ are the query, key, and value vectors and $d_k$ is the dimension of the vector $K$. The softmax function, by definition, is the normalised exponential function that takes an input vector of \textit{K} real numbers and normalises it into a probability distribution consisting of \textit{K} probabilities proportional to the exponential of the input numbers. 
The building blocks of our transformer models are layers applying self-attention and are termed transformer encoders. Please refer to \citet{vaswani2017attention} for a detailed discussion on transformer encoders.

\subsection{LSBG Detection Transformer (LSBG DETR)}
\begin{figure}[h]
\centering
\includegraphics[width=250 pt,keepaspectratio]{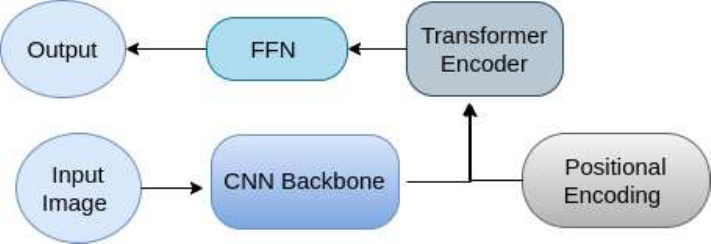}
\caption{Scheme of the general architecture of the Detection Transformer (LSBG DETR) taken from \citet{Hareesh}. The extracted features of the input image by the CNN backbone are combined with positional encoding and passed on to the encoder layer to assign attention scores to each feature. The weighted features are then passed to the feed-forward neural network (FFN) to predict the probability.}
\label{fig:Transformer}
\end{figure}

We implemented four transformer models that use a CNN backbone and self-attention layers to classify the labels, which we call LSBG DETR (LSBG detection transformers) models in general. The LSBG DETR architecture is inspired by transformer models from \citet{Hareesh}, which explored diverse structures and hyperparameters to optimize classification performance.   Each individual model is followed by a number indicating their chronological order of creation. The LSBG DETR models have an 8-layer CNN backbone to extract feature maps from the input image. The feature maps produced by the CNN backbone are then passed on to the transformer encoder layer to create an attention map that helps the transformer component focus on the most relevant features for classification. The transformer encoder layer has sub-components known as heads which parallelly apply the self-attention to the input vector split into smaller parts. Output generated by the transformer encoder is then passed on to a feed-forward neural network (FFN) layer to predict the probability of the input being an LSBG or not. Another point to be noted is that the transformers are permutation invariant; hence, we add positional encoding to address this issue and retain the positional information of features. For the LSBG DETR, we used fixed positional encoding defined by the function 
 \begin{gather}
     PE_{(pos,2i)} = \rm sin \Big(pos/12800^\frac{2i}{d_{model}}\Big),\\
     PE_{(pos,2i+1)} = \rm cos \Big(pos/12800^\frac{2i}{d_{model}}\Big),
 \end{gather}
 where $pos$ is the position, $i$ is the dimension of the positional encoding vector, and $d_{model}$ is the dimension of the input feature vector. We follow the positional encoding defined in \citet{vaswani2017attention}, and for a detailed discussion on positional encoding and its importance, we refer to \citet{liutkus2021relative,su2021roformer,chen2021demystifying}.
 The general structure of the LSBG DETR is shown in Fig.  \ref{fig:Transformer}.
 For a detailed discussion on the transformer models similar to LSBG DETR, we refer to \citet{carion2020endtoend} and \citet{Hareesh}. 

\subsection{LSBG Vision}

\begin{figure*}[b]
\centering
\includegraphics[width=500 pt,keepaspectratio]{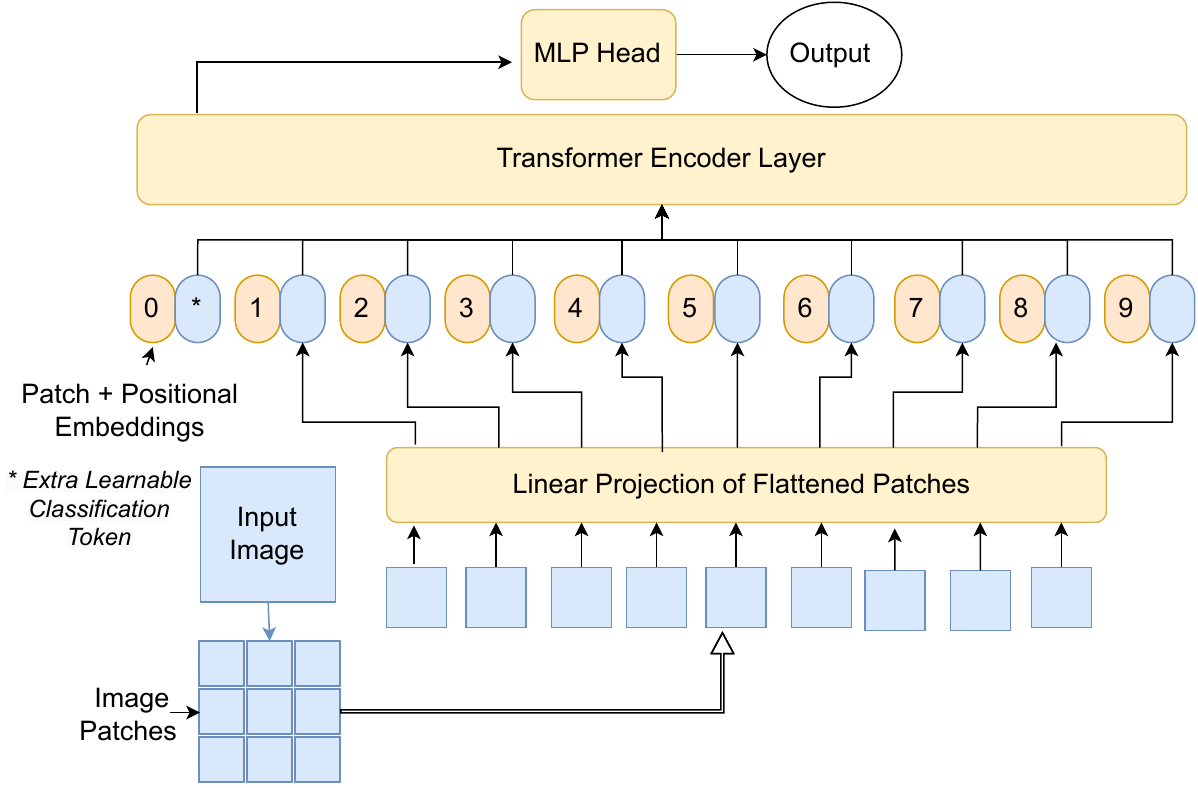}
\caption{Scheme of the general architecture of the LSBG vision transformer (LSBG ViT). The input image is split into small patches and flattened into a sequence of 1D vectors and combined with positional encoding. The numbered circular patches represent the position encoding, and the counterpart represents the flattened 1-D sequence of the image patches. The combined 1-D sequence is passed to the transformer layers. The extra learnable class embedding has encoded the class of the input image after being updated by self-attention and passed it on to an MLP head to predict the output.}
\label{fig:Vision_transformer}
\end{figure*}

We have created four transformer models similar to the Vision Transformer (ViT) introduced by Google Brain \citep{dosovitskiy2020image}, which we call LSBG vision transformers (LSBG ViT) in general. Similar to LSBG DETR models, each individual model is followed by a number indicating their chronological order of creation. One of the main features of LSBG ViT models is that it does not use any convolutional layers to process the image, unlike LSBG DETR. In the ViT architecture, the input image is divided into fixed-size patches, which are flattened into a sequence of 1D vectors. Since the transformers are permutation invariant, the positional embedding is added to the patch embedding before they are fed into the transformer layers. The positional embedding is typically a fixed-length vector that is added to the patch embedding, and it is learned during training along with the other model parameters. The combined 1-D sequence is then passed through a stack of transformer layers. An additional learnable (class) embedding is affixed to the input sequence, which encodes the class of the input image. This class embedding for each input is calculated by applying self-attention to positionally embedded image patches. Output from the class embedding is passed on to a multi-layer perceptron (MLP) head to predict the output class. A schematic diagram of the Vision transformer is shown in Fig.  \ref{fig:Vision_transformer}. For a detailed discussion on ViT models, please refer to \citet{dosovitskiy2020image}.

 \subsection{Training}
 All of the LSBG DETR and LSBG ViT models were trained with an initial learning rate of $\alpha= 10^{-4}$. We used the exponential linear unit (ELU) function as the activation function for all the layers in these models \citep{Clevert}. We initialise the weights of our model with the Xavier uniform initialiser \citep{Glorot}, and all layers are trained from scratch by the ADAM optimiser with the default exponential decay rates \citep{kingma2017adam}. We have used the early stopping callback from {\tt Keras} \footnote{\url{https://keras.io/api/callbacks}} to monitor the validation loss of the model and stop training once the loss was converged. The models LSBG DETR 1 and 4 had 8 heads and were trained for 150 and 93 epochs, respectively. Similarly, the LSBG DETR 2 and 3 had 12 heads and were trained for 134 and 105 epochs, respectively. Coming to the LSBGS ViT models, the hyperparameters we varied were the size of the image patches, the number of heads and the number of transformer encoder layers. The hyperparameters for the all the LSBG DETR models were customized based on the results from \citet{Hareesh}, which extensively investigated the hyperparameter configurations of DETR models. When it comes to the LSBG vision transformer models, we maintained the hyperparameters from the LSBG DETR models such as learning rate, and batch size, except for adjustments in image patch size, the count of attention heads, and the number of transformer encoder layers. We varied these parameters and the four best models are presented in Table \ref{table:1}. In the spirit of
reproducible research, our code for LSBG DETR  and LSBG ViT is publicly available \footnote{\url{https://github.com/hareesht23/}}.

\begin{table}[h]
\centering
\addtolength{\tabcolsep}{10pt}
\caption{Table showing the name of the model, size of the image patches (s), number of heads (h), number of transformer encoder layers (T) and the number of epochs taken to train (e) the four vision models in chronological order of creation.}
\begin{tabular}{c c c c c}
\hline
\textbf{Model Name} & \textbf{s} & \textbf{h} & \textbf{T} & \textbf{e} \\
\hline
LSBG VISION 1 & 4 & 12 & 4 & 55 \\
LSBG VISION 2 & 4 & 12 & 8 & 55 \\
LSBG VISION 3 & 6 & 12 & 4 & 67 \\
LSBG VISION 4 & 6 & 16 & 8 & 67 \\
\hline
\end{tabular}
\label{table:1}
\end{table}

\subsection{Ensemble Models}
We had two classes of transformers (LSBG DETR and LSBG ViT) with four models in each class, and we used an ensemble model of these four models for each class to look for new LSBGs from DES DR1. Ensemble models in deep learning refer to combining multiple models to create a single model that performs better than the individual models. The idea behind ensemble models is to reduce the generalisation error and increase the stability of the system by taking into account multiple sources of information. Various kinds of ensemble learning exist in the literature, and they have been found helpful in a broad range of machine learning problems \citep{wang2022wisdom}. For a detailed review of ensemble methods, please refer to \citet{domingos1999metacost} and \citet{dietterich2000ensemble}. 
One of the easiest and most common ensemble methods is model averaging. In model averaging, multiple models are trained independently on the same training data, and the outputs of the models are averaged to make the final prediction. One of the main advantages of model averaging is that it is computationally efficient and does not require any additional training time. It also allows the use of different types of model architectures and can take advantage of their strengths and weaknesses and improve overall performance. Here we use averaging to create the ensemble models for LSBG DETR and LSBG ViT. 

\subsection{S\'ersic fitting}\label{sect:morphological_fitting}
The candidates identified independently by both LSBG DETR and LSBT ViT ensemble models were subjected to a single component S\'ersic fitting using {\tt Galfit} \citep{Galfit}. This was done to re-estimate the $\bar{\mu}_{eff}$ and $r_{1/2}$ values of our LSBG candidates that we initially used for our sample selection. We employed a single-component S\'ersic fitting method to align with the LSBG search methodology of \citet{Tanoglidis1}, who also utilized a similar approach. However, we also note that S\'ersic fitting does not always capture the full light from a galaxy.

We used the magnitude ({\tt MAG\_AUTO}) and radius ({\tt FLUX\_RADIUS}) values from the gold catalogue as an initial guess for the {\tt Galfit} procedure. Moreover, the S\'ersic index ($n$) and axis ratio ($q$) were initialised to be at a fixed value of 1 and were allowed to vary only within the range of $0.2 < n < 4.0$ and $0.3 < q \leq 1.0$, respectively. A similar fitting procedure was done for both the \textit{g}-band and \textit{i}-band images of our sample. After the fitting, we excluded all the sources with poor/failed fits with either a reduced $\chi^{2}>3$ or if their {\tt Galfit} magnitude estimates diverge from their initial {\tt MAG\_AUTO} values by more than one mag. 
We have also excluded the cases where the estimated $n$ and $q$ values do not converge and are on the edge of the range specified above.
For the remaining galaxies, we re-applied our $g$-band sample selection criteria of $\bar{\mu}_{eff}>24.2$ mag arcsec$^{-2}$ and $r_{1/2}>2.5 {\arcsec}$, following \citet{Tanoglidis1}.
The $\bar{\mu}_{eff}$ values were calculated using the relation given by Eq. \ref{eqn:mu_eff}:
\begin{equation}\label{eqn:mu_eff}
\centering
    \bar{\mu}_{eff} = m + 2.5 \times log_{10}(2\pi r_{1/2}^2),
\end{equation}
where $\bar{\mu}_{eff}$ is the mean surface brightness within the effective radius, \textit{m} is the total magnitude and $r_{1/2}$ is the half-light radius in a specific band estimated from {\tt Galfit}. For all our measurements, we also applied a foreground Galactic extinction correction using the \citet{Schlegel1998} maps normalised by \citet{Schlafly2011} and a \citet{Fitzpatrick1999} dust extinction law.

\subsection{Visual Inspection}
Only the candidates identified independently by LSBG DETR and LSBT ViT ensemble models and passed the selection criteria for being an LSBG with the updated parameters from the {\tt Galfit} were considered for visual inspection. This refined sample was subjected to visual inspection by two authors independently. Candidates identified as LSBG by both authors were treated as confident LSBGs, and candidates identified as LSBG by only one author were reinspected together to make a decision. Since visual inspection is time-consuming, we only resorted to it at the last step and tried as much to reduce the number of candidates shortlisted for visual inspection. 

To aid in visual inspection, we used two images for every candidate. We generated images enhancing the low surface brightness features using the  {\tt APLpy} package \citep{Robitaille} and images downloaded from the DESI Legacy Imaging Surveys Sky Viewer \citep{DESI}. Furthermore, the \textit{g}-band S\'ersic models from {\tt Galfit} were also used to visually inspect the quality of the model fitting. Each candidate was then categorised into three classes based on the {\tt Galfit} model fit and the images: LSBG, non-LSBG (Artifacts), or misfitted LSBGs. If the model of the galaxy was fitted correctly and the candidate showed LSBG features, it was classified as an LSBG. If the candidate shows LSBG features but does not fit correctly, we classify it as a misfitted LSBG. Finally, if the candidate does not have features of an LSBG, we classify it as an artefact or non-LSBG. 

\subsection{Metrics for comparing models}
Here, we use accuracy, true positive rate (TPR), false positive rate (FPR) and area under the receiver operating characteristic (AUROC) curve as the metrics to compare the performance of the created transformer models. The classification accuracy of a model is defined as:
\begin{equation}
    \text{Accuracy} = \frac{TP+TN}{TP+FP+TN+FN},
\end{equation}
where TP is the number of true positives, TN is the number of true negatives, FP is the number of false positives, and FN is the number of false negatives. Since identifying LSBGs with less contamination is our primary focus, rather than the overall accuracy of the classifier, TPR and FPR are more informative metrics for evaluating the classifier's performance. The TPR is the ratio of LSBGs identified by the model to the total number of LSBGs, which can be expressed as
\begin{equation}
    TPR = \frac{TP}{TP + FN}.
\end{equation}
In the literature, sensitivity is another term used to represent the True Positive Rate (TPR), and it measures how well a classifier detects positive instances (in this case, LSBGs) from the total number of actual positive instances in a dataset. Similarly, FPR can be considered a contamination rate because it measures how often the classifier incorrectly classifies negative instances as positive. FPR is defined as
\begin{equation}
    FPR = \frac{FP} {FP + TN}.
\end{equation} 
All the quantities defined above are threshold dependent and vary as a function of the chosen probability threshold. By constructing the receiver operating characteristic curve (ROC) and finding the AUROC, one could define a threshold-independent metric for comparing the models. The ROC curve is constructed by plotting the true positive rate (TPR) and false positive rate as a function of the threshold. The area under the ROC curve (AUROC) measures how well a classifier distinguishes between classes and is a constant for the model, unlike the accuracy, which varies with a threshold. If the AUROC is 1.0, the classifier is perfect with TPR = 1.0 and FPR = 0.0 at all thresholds. A random classifier has an AUROC $\sim$ 0.5, with TPR almost equal to FPR for all thresholds.

\section{Results}\label{results}

\subsection{Model performance on the testing set} 

We have created four models of each transformer, namely LSBG DETR and LSBG ViT, with different hyperparameters to generalise our results for both transformers. Each model was implemented as a regression model to predict the probability of an input being an LSBG, and we set 0.5 as the threshold probability for classifying an input as LSBG. Further, we use an ensemble of the four models as the final model for LSBG DETR and LSBG ViT. Table \ref{table:1} describes the architecture, accuracy and  AUROC of all the models, including the ensemble models on the test dataset, as mentioned in section \ref{traing_data}.
\begin{table}[h]
\centering 
\addtolength{\tabcolsep}{-03.50pt}
\caption{Table comprising the architecture, accuracy, true positive rate (TPR), false positive rate (FPR) and AUROC of all the models in chronological order of creation.}
\begin{tabular}{c c c c c}\hline\hline
\textbf{Model   name}  &\textbf{Accuracy (\%)} &\textbf{TPR} & \textbf{FPR}& \textbf{AUROC} \\\hline
LSBG VISION 1 & 93.55 & 0.97 & 0.12 &  0.980   \\
LSBG VISION 2 & 93.79 & 0.97 & 0.11 & 0.980 \\
LSBG VISION 3 & 93.47 & 0.97 & 0.11 & 0.981\\
LSBG VISION 4 & 93.51 & 0.97 & 0.11 &  0.980 \\
LSBG VISION Ensemble & \textbf{93.75} & \textbf{0.97} & \textbf{0.11} & \textbf{0.983} \\
LSBG DETR 1 & 94.36 & 0.97 & 0.09  & 0.982  \\
LSBG DETR 2 & 94.28 &0.96  &0.08  &  0.980  \\
LSBG DETR 3 & 94.36 & 0.96 & 0.08 &  0.982   \\
LSBG DETR 4 & 94.24 & 0.95 & 0.07 & 0.982 \\
LSBG DETR Ensemble & \textbf{94.60} & \textbf{0.96} & \textbf{0.07} &\textbf{0.984} \\
 \hline
\end{tabular}
\label{table:2}
\end{table}

\begin{figure}[h]
\centering
\includegraphics[width=250 pt,keepaspectratio]{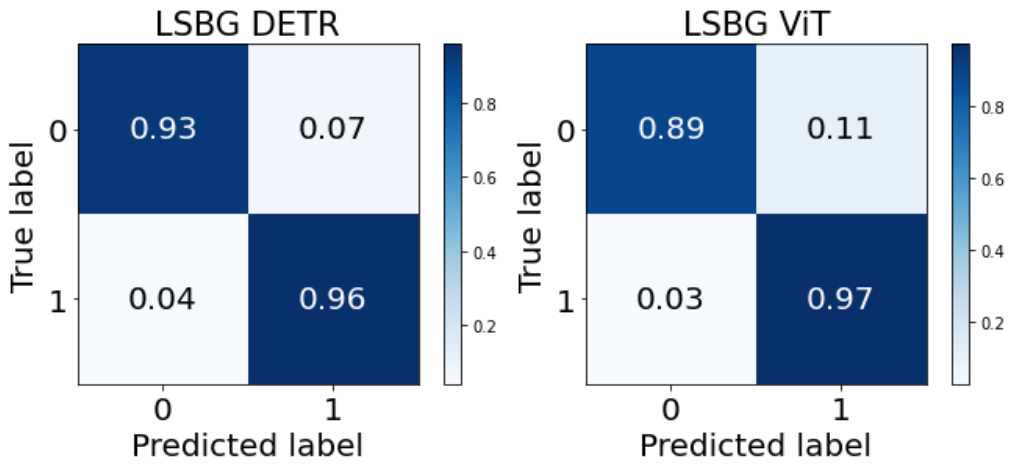}
\caption{Confusion matrix of LSBG DETR and LSBG ViT models plotted for a threshold = 0.5. Class 0 represents the artefacts, and Class 1 represents the low surface brightness galaxies.}
\label{fig:LSBG_CM}
\end{figure}

As mentioned earlier, the more insightful metrics are the true positives (TPR) and the false positives (FRP) rather than overall accuracy. These metrics can be visualised using a confusion matrix, which is shown in Fig.  \ref{fig:LSBG_CM}, for the ensemble models using a threshold of 0.5.
 The LSBG DETR ensemble had a TPR of 0.96 and an FPR of 0.07, indicating that the LSBG DETR ensemble model can accurately identify 96\% of all LSBGs in the DES data, with an estimated 7\% contamination rate in the predicted sample. Similarly, the LSBG ViT Ensemble model can identify 97\% of all the LSBGs in DES but with 11\% contamination. 

\begin{figure}[h]
\centering
\includegraphics[width=250 pt,keepaspectratio]{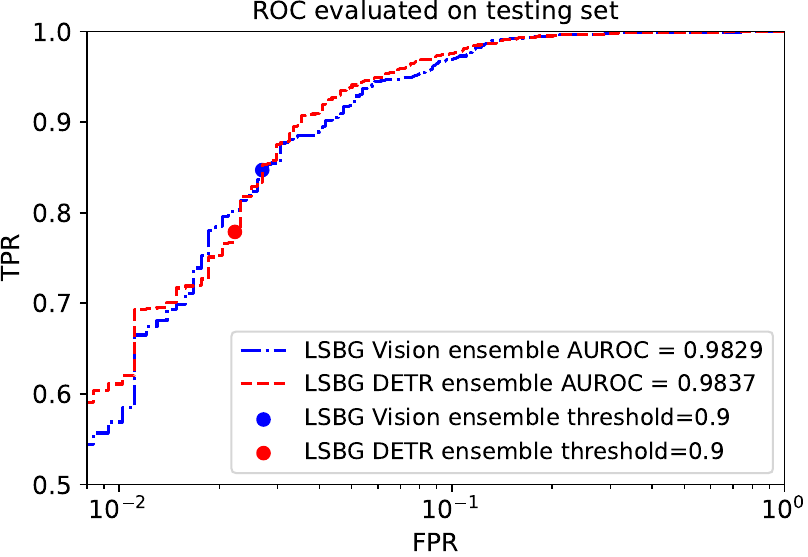}
\caption{Receiver operating characteristic (ROC) curve of the ensemble models. The red and blue lines represent the variation of FPR and TPR as a function of the threshold for LSBG DETR and LSBG Vision ensembles, respectively. The red and blue points mark the TPR and FPR for a threshold = 0.9. }
\label{fig:AUROC}
\end{figure}

 The receiver operator characteristic (ROC) curve of the LSBG DETR and LSBG ViT ensemble models are shown in Fig.~\ref{fig:AUROC}. In terms of accuracy and AUROC, the LSBG DETR models performed slightly better than the LSBG ViT models. It is clear from Fig.  \ref{fig:AUROC} that both the ensemble models have a TPR $\sim 0.75$ even for a high threshold such as 0.9. Indicating that both the ensemble models can confidently identify around $\sim 75\%$ of all the LSBGs in DES and assign these candidates with a probability greater than 0.9. 
\begin{figure*}[h]
\centering
\includegraphics[width=500 pt,keepaspectratio]{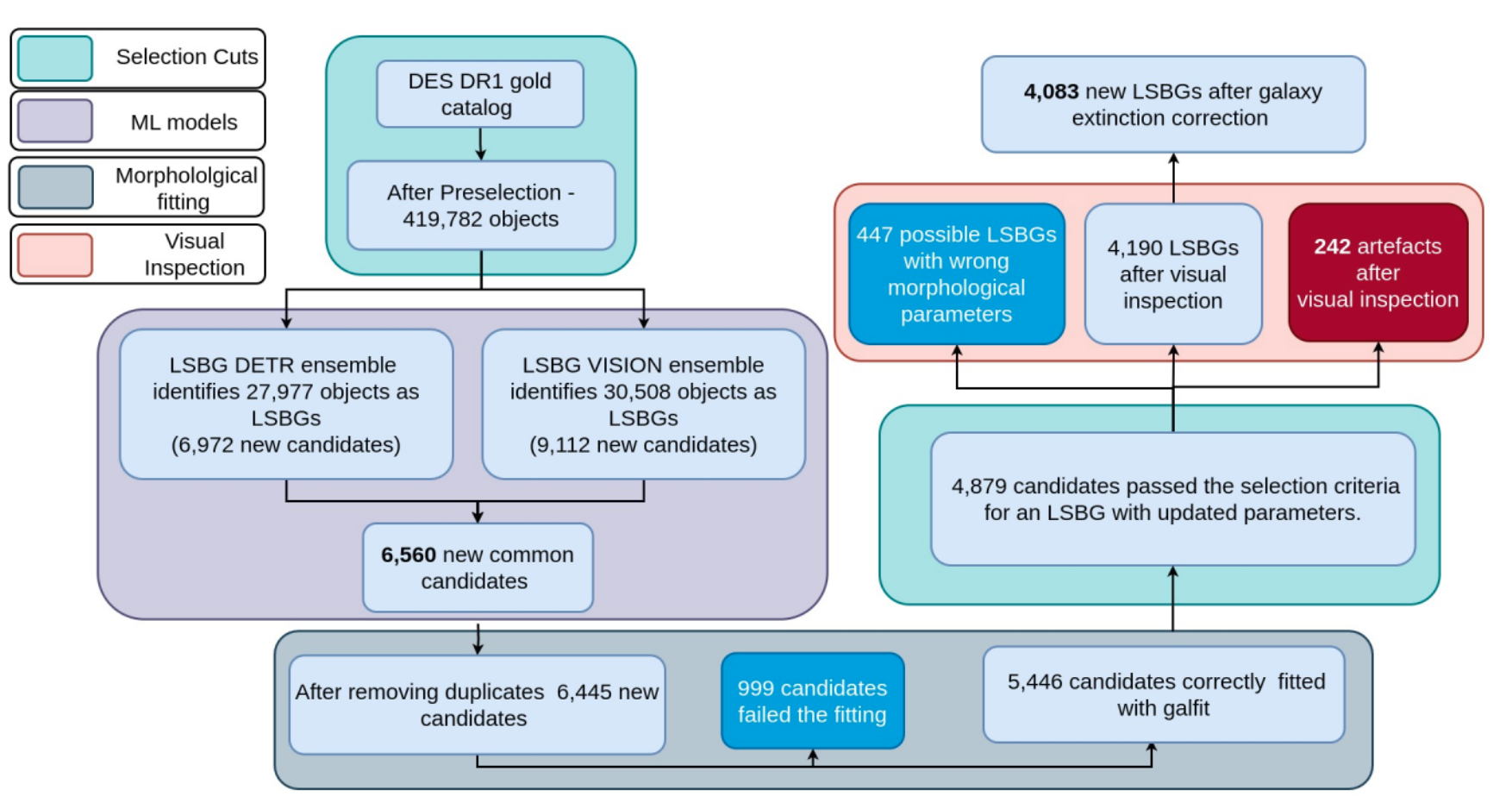}
\caption{Schematic diagram showing the sequential selection steps used to find the new LSBG sample.}
\label{fig:candidate_slection}
\end{figure*}
\subsection{Search for LSBGs in the full coverage of DES}

Since the LSBG DETR model and the LSBG ViT model have different architectures and feature extraction principles, we regard the ensemble models of these two as separate independent transformer classifiers. In order to search for new LSBGs from DES, we employed the transformer ensemble model on the  419\,782 objects that satisfied the selection criteria defined in section \ref{preselection}. The candidates scoring above the threshold probability of 0.5 were catalogued as potential LSBG candidates. The LSBG DETR ensemble classified 27\,977 objects as LSBGs, among which 21\,005 were already identified by \citet{Tanoglidis1}. Similarly, the LSBG ViT ensemble classified 30\,508  objects as LSBGs, among which had 21\,396 LSBGs common with the sample identified by \citet{Tanoglidis1}. So finally, 6\,972 and 9\,112 new candidates were classified as potential LSBGs by the LSBG DETR and LSBG ViT ensembles, respectively. However, only the  6\,560 candidates identified by both the ensemble models independently were considered for further analysis to reduce the false positives. Since there is a possibility that there might be duplicates of the same candidates existing in the selected sample, we ran an automated spatial crossmatch to remove duplicate objects separated by < 5". The origin of these duplicates can be traced back to the fragmentation of larger galaxies into smaller parts by {\tt SourceExtractor}. After removing the duplicates, the number of potential LSBG candidates reduced from 6\,560 to 6\,445. As discussed in Sect.~\ref{sect:morphological_fitting}, these candidates were subjected to single component S\'ersic model fitting using {\tt Galfit}.

\begin{figure*}[h]
\centering
\begin{subfigure}{0.33\textwidth}
  \includegraphics[width=\linewidth]{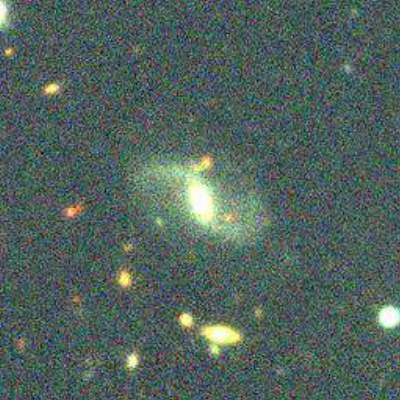}
  \caption{Coadd Object Id - 295747204 }
  \label{fig:lsb1}
\end{subfigure}
\begin{subfigure}{0.33\textwidth}
  \includegraphics[width=\linewidth]{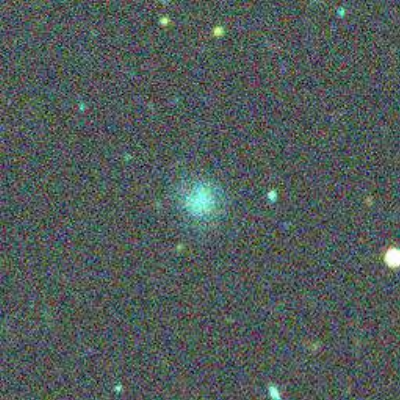}
  \caption{Coadd Object Id - 61515112}
  \label{fig:lsb3}
\end{subfigure}
\begin{subfigure}{0.33\textwidth}
  \includegraphics[width=\linewidth]{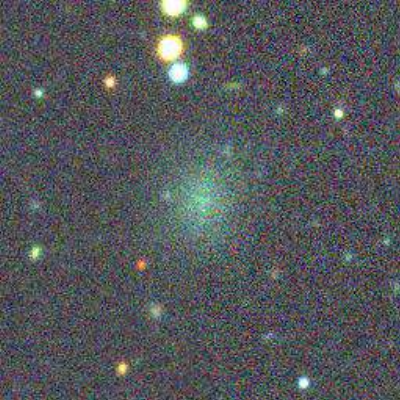}
  \caption{Coadd Object Id - 62646182}
  \label{fig:lsb5}
\end{subfigure}
\begin{subfigure}{0.33\textwidth}
  \includegraphics[width=\linewidth]{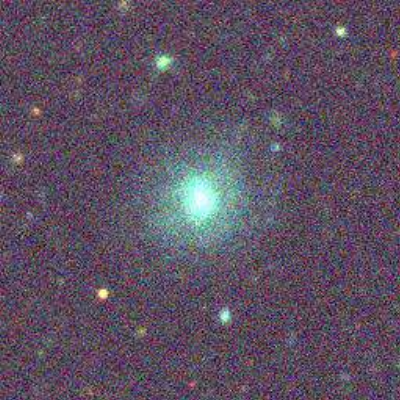}
  \caption{Coadd Object Id - 64560481}
  \label{fig:lsb6}
\end{subfigure}
\begin{subfigure}{0.33\textwidth}
  \includegraphics[width=\linewidth]{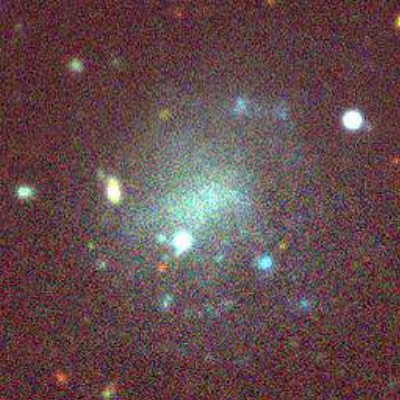}
  \caption{Coadd Object Id - 67813078}
  \label{fig:lsb7}
\end{subfigure}
\begin{subfigure}{0.33\textwidth}
  \includegraphics[width=\linewidth]{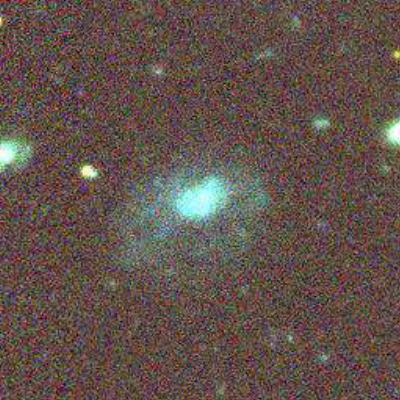}
  \caption{Coadd Object Id - 69253856}
  \label{fig:lsb8}
\end{subfigure}
\begin{subfigure}{0.33\textwidth}
  \includegraphics[width=\linewidth]{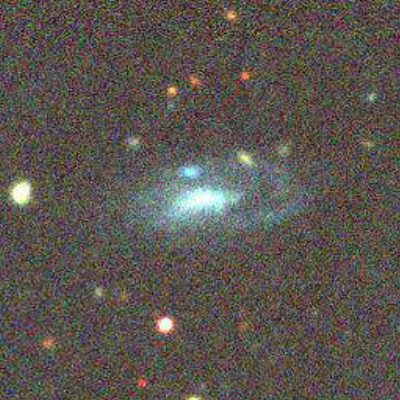}
  \caption{Coadd Object Id - 70739980}
  \label{fig:lsb9}
\end{subfigure}
\begin{subfigure}{0.33\textwidth}
  \includegraphics[width=\linewidth]{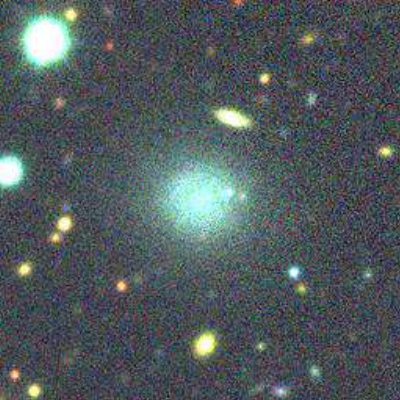}
  \caption{Coadd Object Id - 73726929}
  \label{fig:lsb11}
\end{subfigure}
\begin{subfigure}{0.33\textwidth}
  \includegraphics[width=\linewidth]{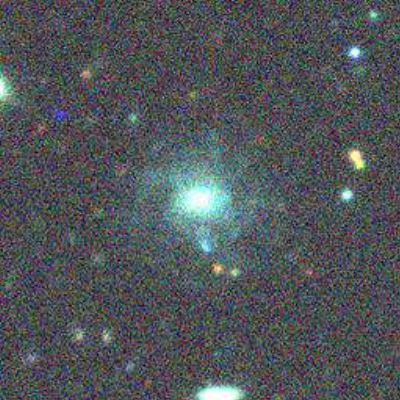}
  \caption{Coadd Object Id - 76917094}
  \label{fig:lsb12}
\end{subfigure}

\caption{Cutouts of 9 confirmed new LSBGs after visual inspection. The unique identification number (co object id) for each galaxy in DES DR1 is given below each image. The images were generated by combining the $g,r$ and $z$ bands using {\tt APLpy} package \citep{Robitaille}, and each image corresponds to a $ 67.32 {\arcsec} \times  67.32 {\arcsec}$ region of the sky with the LSBG at its centre.}
\label{fig:lsbs_vis}
\end{figure*}

During the {\tt Galfit} modelling, 999 candidates had failed fits and were consequently removed from the sample since our objective is to produce a high-purity sample with accurate S\'ersic  parameters. We visually inspected these unsuccessful fits and found that in most cases the presence of a very bright object near the candidate was the cause of the poor S\'ersic fit. Of the remaining 5\,446 candidates, 4\,879 passed the \mue{} and \re{} selection criteria outlined in Sect. \ref{preselection} with the updated parameters. These 4\,879 candidates were inspected visually to identify the genuine LSBGs. After independent visual inspections by the authors,  4\,190 candidates were classified as LSBGs and 242 candidates were found to be non-LSBGs. During visual inspection, 447 candidates were found to be possible LSBGs with unreliable measurements from {\tt Galfit}. These candidates are excluded from our final sample, and here we only report on the most confident candidates that were identified as LSBGs during visual inspection. After correcting for the Galactic extinction correction,
our final sample reduced to 4\,083 new LSBGs from DES DR1. The schematic diagram showing the sequential selection steps used to find the new LSBG sample is shown in Fig. \ref{fig:candidate_slection}. A sample catalogue comprising the properties of the newly identified LSBGs is shown in table \ref{table:newsample}, and some examples of the new LSBGs that we have found are plotted in Fig. \ref{fig:lsbs_vis}.

\begin{sidewaystable*}
\caption{Sample of new LSBGs identified in this work.} 
\addtolength{\tabcolsep}{-.50pt}
\begin{tabular}{cccccccccccccccc}
\hline
COADD\_ID &      RA &       DEC &   $g_{gf}$ &   $g_{cor}$ &   $\bar{\mu}_{g\_eff\_gf}$ &  $r_{g 1/2}$ &   n &    q &  $log_{10}(\Sigma_{star})$ & $\chi_{\nu g}^2$ &   $i_{gf}$ &   $i_{cor}$ &   $\bar{\mu}_{i\_eff\_gf}$ &  $r_{i, 1/2}$ & $\chi_{\nu i}^2$  \\
 & (deg) & (deg) &    &  &   (\small{mag arcsec$^{-2}$}) &   (arcsec)&   &   &  (M$_{\odot}$ kpc$^{-2}$) & &   &   &   (\small{mag arcsec$^{-2}$}) &  (arcsec) & \\
\hline
          61456395 & 29.7062 & -60.4882   &      19.06 &       18.97 &                          25.41 &          9.41 & 2.17 & 0.62 &                        5.86 &               1.06 &      18.72 &       18.67 &                          25.37 &         10.86 &               1.03 \\
          61508029 & 29.925  &   4.60483  &      19.84 &       19.67 &                          24.33 &          3.43 & 0.72 & 0.85 &                        6.82 &               1.03 &      19.27 &       19.19 &                          23.52 &          3.06 &               0.99 \\
          61580602 & 30.3125 & -58.1927   &      18.89 &       18.83 &                          24.56 &          5.99 & 1.5  & 0.82 &                        6.87 &               1.02 &      18.07 &       18.04 &                          23.72 &          5.93 &               1.02 \\
          61638403 & 29.9539 &   4.75811  &      20.02 &       19.87 &                          24.28 &          3.84 & 0.74 & 0.54 &                        6.52 &               1.04 &      19.61 &       19.53 &                          23.62 &          3.45 &               1    \\
          61638933 & 29.4862 &   4.74468  &      19.2  &       19.06 &                          24.24 &          4.59 & 0.93 & 0.79 &                        6.82 &               0.98 &      18.7  &       18.63 &                          23.38 &          3.89 &               0.99 \\
          61712539 & 30.2824 &   5.28992  &      19.67 &       19.53 &                          24.26 &          4.57 & 1.01 & 0.52 &                        6.58 &               1.01 &      19.47 &       19.4  &                          23.2  &          3.07 &               1.03 \\
          61766250 & 30.1112 &  -9.24769  &      20.76 &       20.69 &                          26.29 &          7.8  & 1.47 & 0.42 &                        5.87 &               1.13 &      20.19 &       20.15 &                          25.2  &          6.15 &               1.1  \\
          62011325 & 29.9449 &  -6.78979  &      21.6  &       21.51 &                          25.91 &          3.17 & 0.5  & 0.84 &                        6.27 &               1.02 &      20.8  &       20.76 &                          25.21 &          3.31 &               1    \\
          62053525 & 29.8595 & -16.8586   &      19.87 &       19.8  &                          24.24 &          3.49 & 0.87 & 0.73 &                        6.81 &               1.03 &      19.32 &       19.29 &                          23.41 &          3.06 &               1.02 \\
          62071354 & 29.8342 & -17.1604   &      20.53 &       20.46 &                          24.47 &          3.01 & 0.84 & 0.66 &                        6.71 &               0.99 &      19.91 &       19.87 &                          23.65 &          2.75 &               0.99 \\
          62227622 & 29.5063 & -13.3511   &      20.84 &       20.79 &                          24.88 &          2.68 & 0.68 & 0.91 &                        7.24 &               1.02 &      19.97 &       19.94 &                          22.96 &          1.66 &               1.05 \\
          62371677 & 29.6675 &  -5.35506  &      19.7  &       19.62 &                          24.25 &          3.54 & 0.79 & 0.84 &                        6.97 &               1.05 &      19    &       18.96 &                          23.32 &          3.2  &               1.03 \\
          62646182 & 30.4643 & -24.0018   &      18.46 &       18.4  &                          24.44 &          6.88 & 1.19 & 0.83 &                        6.77 &               1    &      17.98 &       17.95 &                          23.59 &          5.8  &               0.97 \\
          62830903 & 29.9594 & -28.8656   &      20.11 &       20.07 &                          24.35 &          2.94 & 0.71 & 0.91 &                        7.56 &               1.03 &      19.01 &       18.99 &                          22.44 &          2.03 &               1.04 \\
          62840965 & 30.1905 &  -6.18018  &      20.18 &       20.1  &                          24.35 &          2.74 & 0.76 & 0.98 &                        7.35 &               1.01 &      19.22 &       19.18 &                          22.85 &          2.14 &               1    \\
          63037584 & 28.9741 & -59.5261   &      19.46 &       19.39 &                          24.4  &          6.33 & 0.9  & 0.38 &                        6.31 &               1.01 &      18.88 &       18.85 &                          23.99 &          6.81 &               1.01 \\
          63097874 & 29.3755 & -61.1452   &      20.6  &       20.51 &                          24.87 &          3.08 & 1.02 & 0.85 &                        6.58 &               1.02 &      20.29 &       20.25 &                          23.86 &          2.23 &               1.02 \\
          63113174 & 29.1535 & -61.4326   &      18.81 &       18.72 &                          24.78 &          6.73 & 0.8  & 0.86 &                        6.47 &               1    &      18.52 &       18.47 &                          24.15 &          5.76 &               0.99 \\
          63262376 & 30.2929 & -24.9787   &      21.48 &       21.43 &                          26.1  &          3.92 & 0.98 & 0.73 &                        6.18 &               1.03 &      20.62 &       20.6  &                          25.35 &          4.12 &               1.04 \\
          63527438 & 30.1134 &  -4.56234  &      18.8  &       18.73 &                          24.51 &          6.72 & 0.42 & 0.68 &                        6.2  &               1.02 &      18.68 &       18.65 &                          24.37 &          6.63 &               0.99 \\
          63716543 & 30.0657 & -39.5446   &      20.58 &       20.52 &                          25.1  &          3.73 & 1.16 & 0.74 &                        6.67 &               1.02 &      19.84 &       19.81 &                          24.01 &          3.17 &               1.01 \\
          63922768 & 29.5244 & -32.7958   &      18.41 &       18.35 &                          24.6  &          7.62 & 1.61 & 0.83 &                        6.59 &               1.08 &      18.06 &       18.03 &                          23.87 &          6.37 &               1.04 \\
          64480503 & 29.4619 & -23.0107   &      18.98 &       18.94 &                          24.33 &          5.05 & 0.91 & 0.86 &                        6.81 &               1.04 &      18.5  &       18.48 &                          23.54 &          4.39 &               1    \\
          64560481 & 30.0521 &  -5.09785  &      18.18 &       18.1  &                          24.36 &          7.21 & 1.17 & 0.91 &                        6.86 &               1.06 &      17.66 &       17.62 &                          23.49 &          6.14 &               1.01 \\
          64697654 & 29.5818 &  -8.50243  &      18.36 &       18.28 &                          24.25 &          6.46 & 1    & 0.87 &                        6.81 &               0.97 &      17.87 &       17.83 &                          23.53 &          5.78 &               0.96 \\
          64773733 & 29.0012 & -50.2462   &      19.39 &       19.32 &                          24.7  &          5.13 & 1.4  & 0.8  &                        6.66 &               1.03 &      18.94 &       18.91 &                          23.78 &          4.13 &               1.01 \\
          64868340 & 29.3394 & -23.8729   &      19.43 &       19.38 &                          24.2  &          3.72 & 0.86 & 0.93 &                        7.63 &               1.02 &      18.36 &       18.34 &                          22.23 &          2.47 &               1.26 \\
\hline
\end{tabular}

\label{table:newsample}
\tablefoot{'COADD\_ID' is the unique id of the source, and 'RA' and 'DEC' gives the sky coordinates of the source as estimated from DES Y3 gold catalogue \citep{Sevilla}. Columns '$g_{gf}$', '$g_{cor}$', '$\bar{\mu}_{g\_eff\_gf}$',  and '$r_{g 1/2}$' represents the magnitude in $g$ band, the $g$ band magnitude after correcting for Galactic extinction, mean surface brightness and the half-light radius for the $g$-band fitting using {\tt Galfit}, respectively. The columns 'n','q', and  $log_{10}(\Sigma_{star})$ represent the S\`ersic index, axis ratio, and the stellar mass density, respectively. Column '$\chi_{\nu g}^2$' represent the reduced chi-square value for the $g$-band fitting using {\tt Galfit}. Similarly, columns  '$i_{gf}$', $i_{cor}$', '$\bar{\mu}_{i\_eff\_gf}$','$r_{i 1/2}$', and  '$\chi_{\nu i}^2$' represents the  magnitude in $i$ band, the $i$ band magnitude after correcting for Galactic extinction, the mean surface brightness, the half-light radius and the reduced chi-square value for the $i$ band fitted using {\tt Galfit}, respectively.} 
\end{sidewaystable*}

\begin{figure*}[h]
\centering

\begin{subfigure}{0.475\textwidth}
  \includegraphics[width=\linewidth]{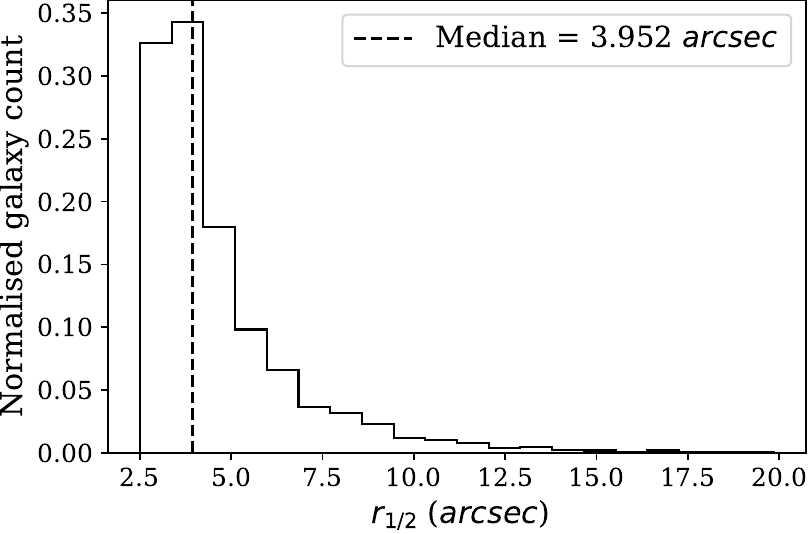}
  \label{fig:lsbr}
\end{subfigure}
\begin{subfigure}{0.475\textwidth}
  \includegraphics[width=\linewidth]{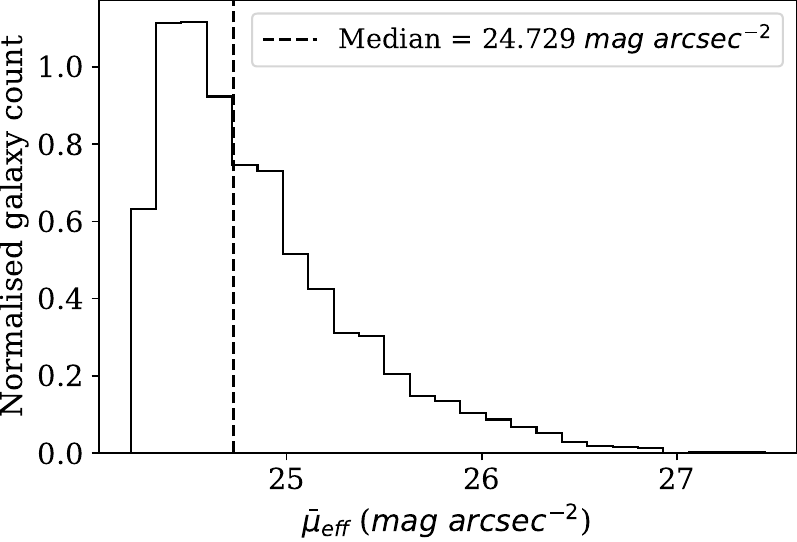}
  \label{fig:lsbmu}
\end{subfigure}
\begin{subfigure}{0.475\textwidth}
  \includegraphics[width=\linewidth]{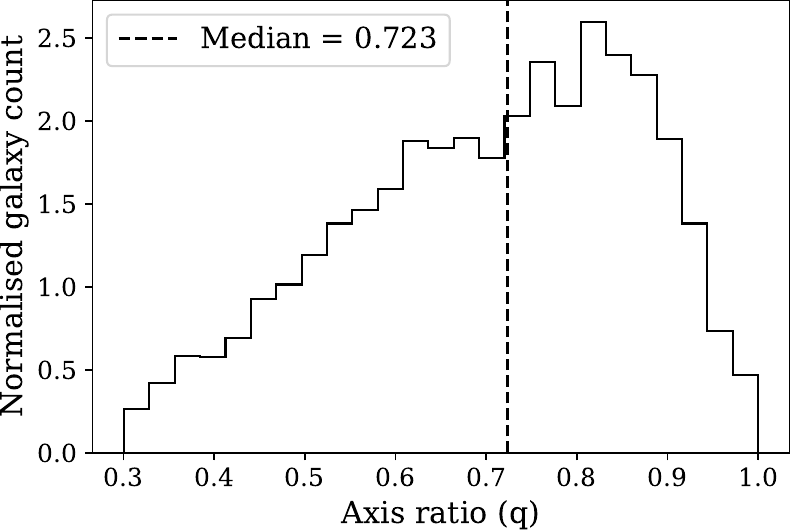}
  \label{fig:lsbq}
\end{subfigure}
\begin{subfigure}{0.475\textwidth}
  \includegraphics[width=\linewidth]{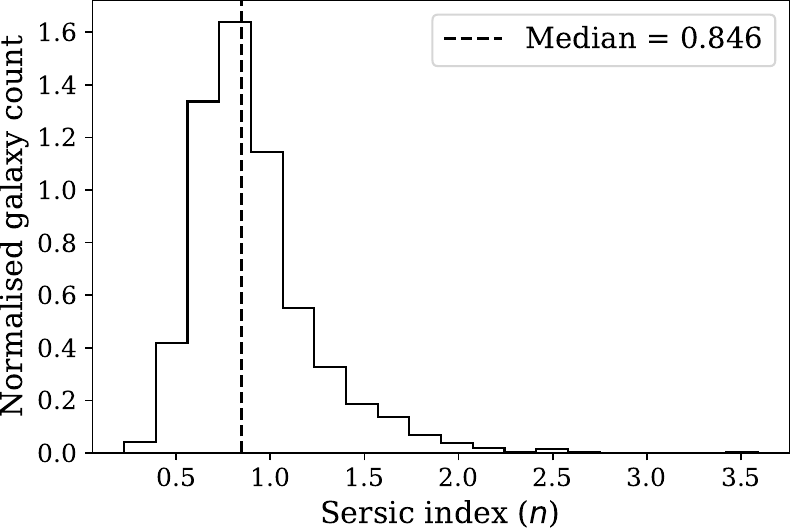}
  \label{fig:lsbn}
\end{subfigure}
\caption{Normalised distribution of half-light radius (top left panel), mean surface brightness (top right panel), S\'ersic index (bottom left panel) and axis ratio (bottom right panel) of the new sample of LSBGs. The dashed line shows the median of the distribution}.\label{fig: lsbs}

\end{figure*}

The distribution of the \re{}, \mue{}, S\'ersic index (\textit{n}), and axis ratio (\textit{q}) of the new sample of LSBGs is plotted in Fig.  \ref{fig: lsbs}. The majority of the LSBGs in this new sample have $r_{1/2} < 7\arcsec$, and \mue{ $< 26$} \magperarcsec{}. 
 The S\'ersic index of the new LSBG sample predominantly lies between 0.5 and 1.5 and has a median value of 0.85. This pattern is similar to the trend identified by \citet{Poulain} in the case of dwarf ellipticals, suggesting that a significant portion of the LSBGs sample could be comprised of such sources. In the case of the axis ratio, the new LSBG sample has a median axis ratio of 0.72 and has a distribution lying in the range of 0.3 to 1. 
 The median value of 0.72 suggests that most galaxies in this sample have a slightly flattened or elongated shape. A detailed discussion of the properties of the new LSBGs identified in this work and their comparison with LSBGs identified by \citet{Tanoglidis1} is presented in Sect. \ref{discs}.
 
\section{Discussion}\label{discs}

\subsection{Transformers as LSBG Detectors}
In this study, we introduce the use of transformers as a classifier model for finding the undiscovered LSBGs in DES. Currently, in the literature, one of the reported deep-learning-based models for classifying LSBGs and artefacts is a CNN model named DeepShadows created by \citet{Tanoglidis2}. They used the catalogue of LSBGs and artefacts identified from DES reported in \citet{Tanoglidis1} to generate the training data. The DeepShadows model achieved an accuracy of 92\% in classifying LSBGs from artefacts and had a true positive rate of 94\% with a threshold of 0.5. Moreover, the DeepShadows model also achieved an AUROC score of 0.974 on this training dataset. However, the DeepShadows was not applied to the complete DES data and checked how it would perform. Nevertheless, DeepShadows was the first deep-learning model to classify LSBGs and artefacts. In addition, \citet{Tanoglidis2} also showed that the DeepShaodws was a better classifier than the support vector machine or random forest models. However, in our work, all of our transformer models were able to surpass the DeepShadows model in every metric individually, which can be seen from Table \ref{table:2}. Namely, in their respective classes, LSBG DETR 1 and LSBG ViT 2 had the highest accuracies ($94.36\% \text{ and } 93.79\%$), respectively.

 Earlier searches for LSBGs have used semi-automated methods such as pipelines based on {\tt imfit} by \citet{Greco} or simple machine-learning models such as SVMs by \citet{Tanoglidis1}. However, the success rate of these methods was very low, and the final candidate sample produced by these methods had around 50\% false positives, which had to be removed by visual inspection. Here we explore the possibilities of transformer architectures in separating LSBGs from artefacts. We used two independent ensemble models of LSBG DETR and LSBG ViT models and single component S\'ersic model fitting to filter the LSBG candidates. Our final sample contained only $\sim 5\%$ as non-LSBGs, which is a significant improvement compared to the previous methods in the literature. Following the definition of an LSBG as described in  \citet{Tanoglidis1}, we identified 4\,083 new LSBGs from DES DR1, increasing the number of identified LSBGs in DES by $17\%$. Our results highlight the significant advantage of using deep-learning techniques to search for LSBGs in the upcoming large-scale surveys. 

\begin{figure*}[t]
\centering
\hfill
\begin{subfigure}{0.24\textwidth}
  \includegraphics[width=\linewidth]{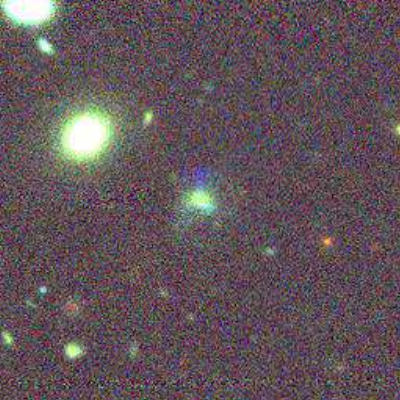}
  \caption{}
  \label{fig:nonlsb1}
\end{subfigure}
\begin{subfigure}{0.24\textwidth}
  \includegraphics[width=\linewidth]{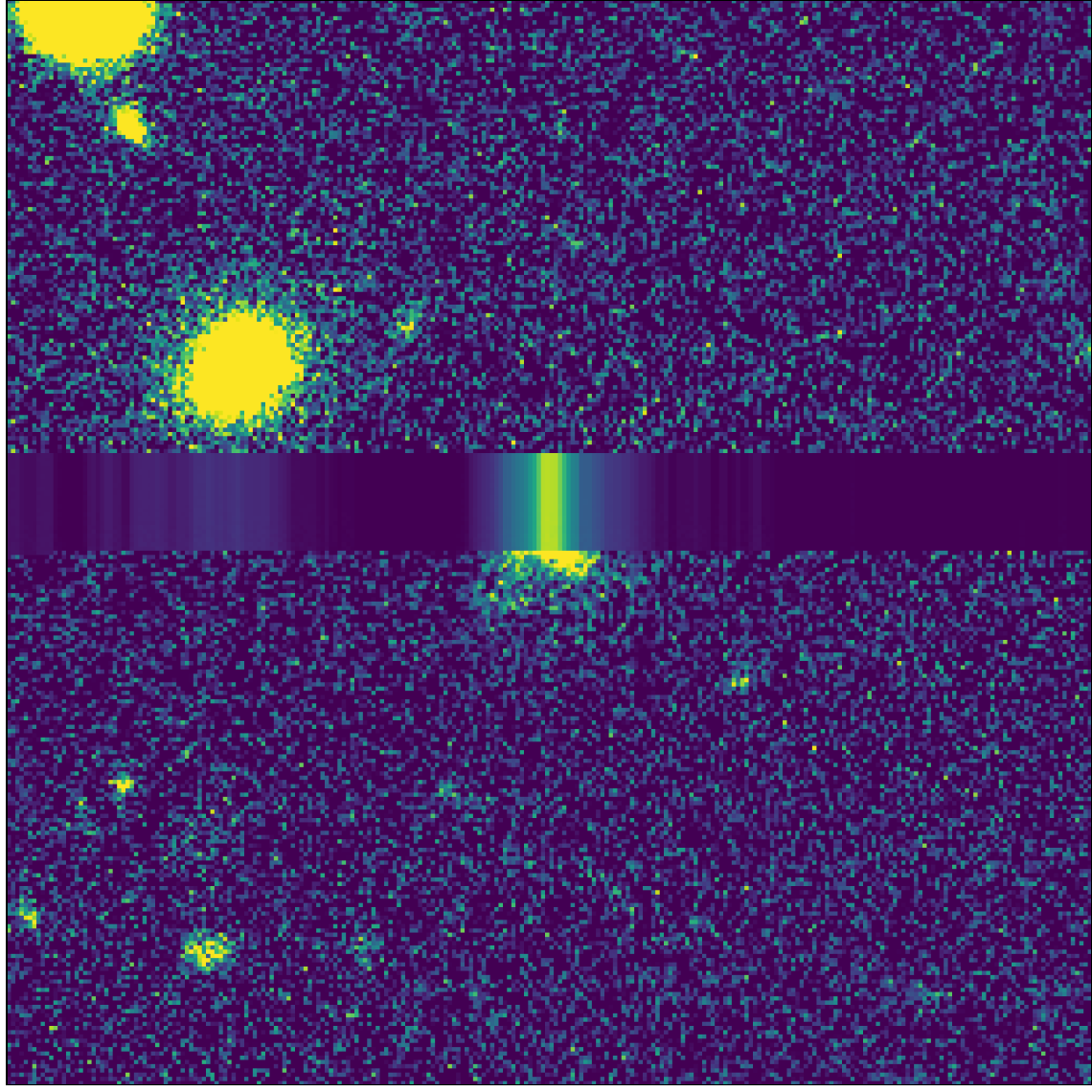}
  \caption{}
  \label{fig:nonlsb2}
\end{subfigure}
\begin{subfigure}{0.24\textwidth}
  \includegraphics[width=\linewidth]{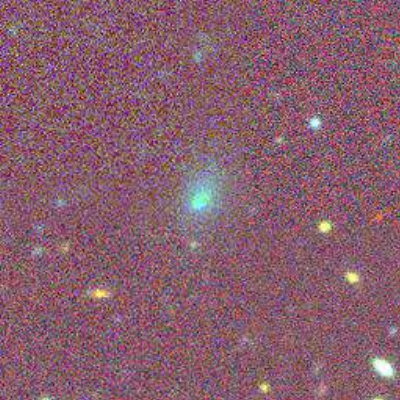}
  \caption{}
  \label{fig:nonlsb3}
\end{subfigure}
\begin{subfigure}{0.24\textwidth}
  \includegraphics[width=\linewidth]{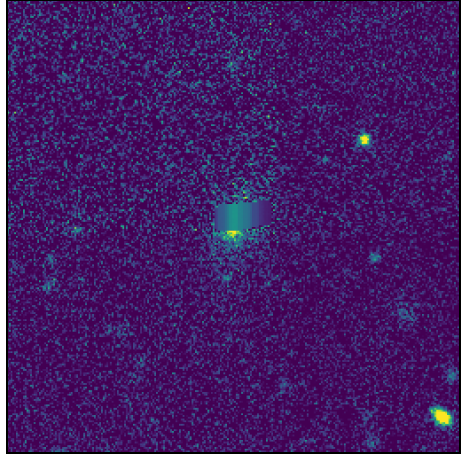}
  \caption{}
  \label{fig:nonlsb4}
\end{subfigure}

\caption{Examples of candidates (Coadd object id - 149796289 and 374192591) classified as non-LSBGs during visual inspection because of glitches in the g-band near the galaxy. The panels (a) and (c) show the RGB image created using the $g,r \text{ and } z$ bands with {\tt APLpy} package \citep{Robitaille}. The panels (b) and (d) show the image in the $g$ band. Each image corresponds to a $ 67.32 {\arcsec} \times  67.32 {\arcsec}$ region of the sky with the candidate at its centre.}
\label{fig:glitches}

\end{figure*}

\begin{figure}[h]
\centering
\hfill
\begin{subfigure}{0.475\textwidth}
  \includegraphics[width=\linewidth]{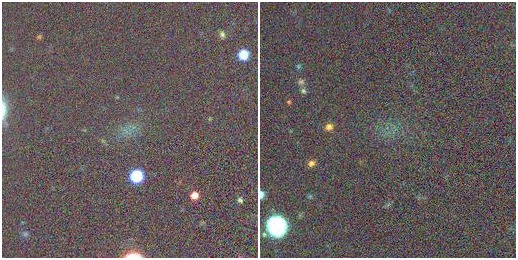}
  \caption{Coadd object id - 251235955 (left) and 99585243 (right)}
  \label{fig:image3}
\end{subfigure}
\hfill
\begin{subfigure}{0.475\textwidth}
  \includegraphics[width=\linewidth]{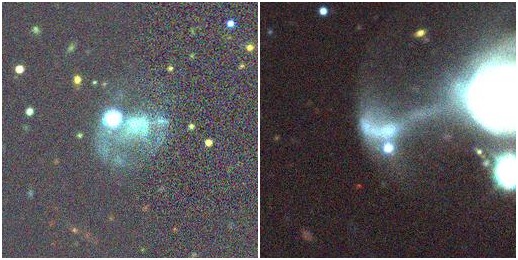}
  \caption{Coadd object id - 125313682 (left) and 113818243 (right)}
  \label{fig:image2}
\end{subfigure}

\caption{Examples of candidates classified as non-LSBG during visual inspection because of lack of sufficient signal in the g-band (a) are shown in the top panel. Candidates classified as non-LSBG during visual inspection because of being artefacts are shown in the lower panel (b). The RGB images are created using the $g,r \text{ and } z$ bands with {\tt APLpy} package \citep{Robitaille}. Each image corresponds to a $ 67.32 {\arcsec} \times  67.32 {\arcsec}$ region of the sky with the candidate at its centre.}
\label{fig:artefacts_vis}

\end{figure}

To have more insights into the fraction of false positives from our method, we checked the performance of these models during training. We encountered around $7\%$ and $11\%$ of artefacts from the LSBG DETR ensemble and LSBG ViT sample, respectively, during training on the test dataset. However, using a combination of these models, we reduced the artefact fraction to less than 5\% during visual inspection. Most of the non-LSBGs we encountered during the visual inspection were faint compact objects that blended in the diffuse light from nearby bright objects. We use the term 'non-LSBG' instead of artefacts here because, during the visual inspection, we classified some potential LSBGs as non-LSBG; these are objects for which the \textit{g}-band images had instrumental artefacts or lack of sufficient signal in the \textit{g}-band. Since the machine learning model intakes three bands as input ($g,r \text{ and } z)$, this suggests that the model was able to study and generalise the nature of LSBGs in each band and was able to predict if it is an LSBG or not based on the signal from the other bands. However, since we define LSBGs based on their \textit{g}-band surface brightness and radius in this work, we classified the galaxies without reliable \textit{g}-band data as non-LSBGs. Some non-LSBGs we encountered during the visual inspection are shown in Fig. \ref{fig:glitches} and  Fig. \ref{fig:artefacts_vis}. With the upcoming surveys of deeper imaging, these galaxies might be classified as LSBGs which might further reduce the non-LSBGs in our candidate sample. 

\begin{figure*}[h]
\begin{subfigure}{0.495\textwidth}
\centering
\includegraphics[width=\linewidth]{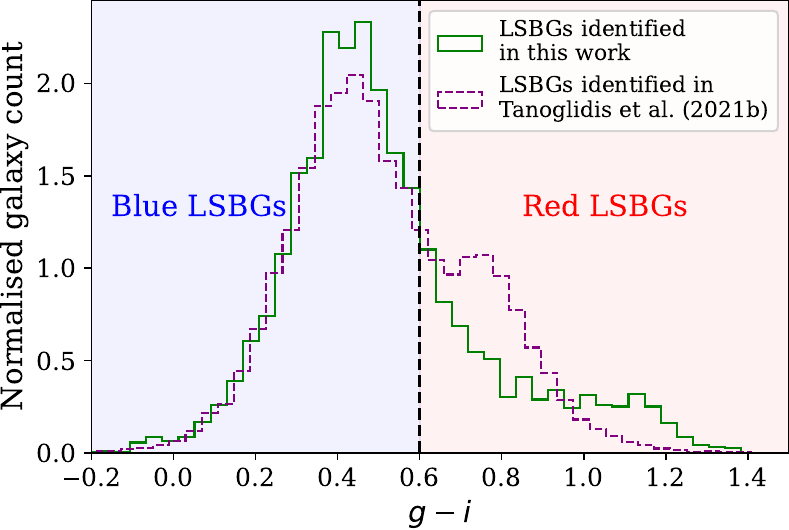}
\label{fig:clr_dis}
\end{subfigure}
\begin{subfigure}{0.495\textwidth}
\centering
  \includegraphics[width=\linewidth]{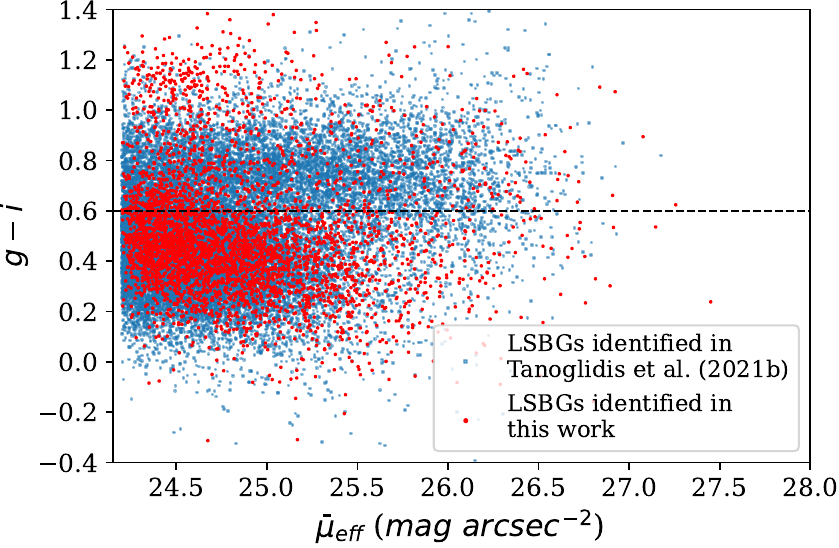}
  \label{fig:clr_mues}
\end{subfigure}
\caption{Normalised distribution of color of the LSBGs from the new sample identified in this work and the LSBGs identified by \citet{Tanoglidis1} plotted in the left panel. The right panel shows the color as a function of mean surface brightness in the $g$-band for the new sample identified in this work and the LSBGs identified by \citet{Tanoglidis1}. The dashed line shows the separation between red and blue LSBGS.}
\label{fig:clr}
\end{figure*}

One another fact to notice when discussing the non-LSBGs from the candidate sample is that some of the candidates identified as LSBGs by the ensemble models (567 out of 5\,446) did not meet the selection criteria of being an LSBG after being fitted with {\tt Galfit}. These galaxies had \re{} ranging from 2\arcsec{} to 20\arcsec{}, with a median of 3.85\arcsec{}, which is similar to the new LSBG sample we found. However, the majority of these galaxies have a mean surface brightness between $24.0-24.2$ \magperarcsec{}, with a median of 24.16 \magperarcsec{}. This suggests that the machine learning model understood the criteria for angular size for LSBGs during its training, but it did not learn the strict condition about the surface brightness. This situation is similar to a human expert analysing a galaxy image to determine whether it is an LSBG or not. Features such as the size of the galaxy are easily identifiable to the human eye. However, determining the surface brightness accurately with only the human eye would be challenging, and there may be possible errors near the threshold region, similar to our machine learning model. So one could say that the machine learning model is behaving approximately like a human visual expert.

Judging from the performance of our model on the training data, we cannot assert that we have discovered all the possible existing LSBGs from the DES DR1. As we can see from Fig.  \ref{fig:LSBG_CM}, the TP rate for the individual ensemble models were 0.96 and 0.97, respectively. This means that the model has not found all the possible LSBGs and a minor fraction of LSBGs is yet to be found in DES DR1. Moreover, to reduce the FPR and the burden during the visual inspection, we have only visually inspected the candidates identified commonly by both the ensemble models and passed the criteria for correctly fitting by {\tt Galfit}.

One of the notable things about the models in this work is that we are using two different ensemble models, each having four models in the ensemble. As we mentioned earlier, each ML model can be considered equivalent to a human inspector, and the ensemble models help balance out the disadvantages of the other models in the ensemble. A closer look at the individual probability distribution of these modes shows that there are 310 candidates among the 4\,083 confirmed LSBG candidates, which had a probability of less than 0.5  for at least one model among the individual models. 
However, since we used an average ensemble model, we were able to identify these LSBG by balancing out the probability, which shows the advantages of using an ensemble model over a single model.

Here, we use the visual inspection as the final step to confirm the authenticity of an LSBG detected by the models. However, it is essential to acknowledge the potential for human bias during the visual inspection, which can impact the accuracy and reliability of the results. For example, during the visual inspection, the visual inspectors disagreed on labelling approximately 10\% of the candidate sample. Most of these galaxies had a mean surface brightness greater than 25.0 \magperarcsec{},  which suggests that even for human experts, it is challenging to characterise extremely faint LSBGs. However, with better imaging, this might change, but we must acknowledge that there will always be some human bias and error associated with human inspection. Also, we must consider that in the upcoming surveys, such as LSST or Euclid, the data size will be too large to inspect visually. In this scenario, relying solely on improved automated methods to purify the sample and accepting a small fraction of false positives could be a feasible solution.

\section{The new sample of DES DR1 LSBGs}\label{newlsbs}

\subsection{The newly identified LSBG sample}
The optical color of a galaxy can give insights into its stellar population. Conventionally, based on their color, the galaxies are divided into red and blue galaxies, and it has been known that color is strongly correlated to the morphology of a galaxy \citep{Strateva}. Blue color galaxies are usually found to be highly active star-forming spiral or irregular systems. In contrast, red color galaxies are mostly found to be spheroidal or elliptical. In addition, the red galaxies have also been found to tend to cluster together compared to the blue galaxies \citep{Bamford}. 

The LSBGs found by \citet{Tanoglidis1} have found a clear bimodality in the $g-i$ color distribution similar to the LSBGs found by \citet{Greco}. In the left panel of Fig. \ref{fig:clr}, we present the $g-i$ color distribution of the 4\,083 new LSBGs and the 23,790 LSBGs found by \citet{Tanoglidis1}. 
We follow the criteria defined by \citet{Tanoglidis1} to define red galaxies as galaxies having $g-i>0.6$ and blue galaxies as galaxies having $g-i<0.6$ where $g$ and $i$ represent the magnitude in each band. In the right panel of Fig. \ref{fig:clr}, we present the color as a function of mean surface brightness in $g$-band for the new sample identified in this work and the LSBGs identified by \citet{Tanoglidis1}. There are 1112 red LSBGs and 2,944 blue LSBGs in the new LSBG sample. \footnote{27 LSBGs failed the modelling using {\tt Galfit} for $i$-band, and they are not included in this color analysis.} From Fig, \ref{fig:clr}, we can see that we have identified a relatively large fraction of blue LSBGs compared to \citet{Tanoglidis1} and a considerable fraction of new red LSBGs with $g-i \geq 0.80$ and a mean surface brightness less than 25.0 \magperarcsec{}. The bias against blue LSBGs and highly red LSBGs in the sample created by \citet{Tanoglidis1} may have been caused by the bias in the training set used to create the SVM, which preselected the LSBG candidates. This bias could have occurred because a large fraction of their training set consisted of LSBGs near the Fornax cluster, which are mainly red LSBGs.

\begin{figure}[ht]
\centering
  \includegraphics[width=\linewidth]{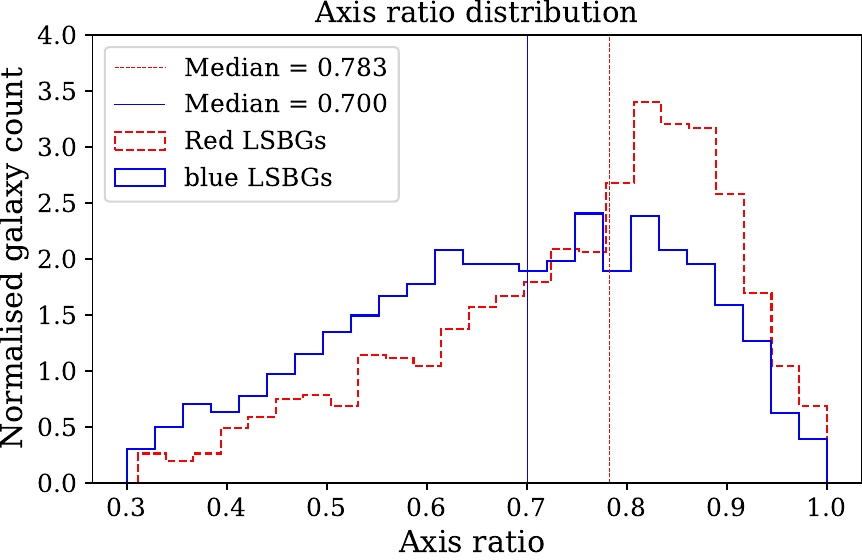}
\caption{Normalised distribution of axis ratio (left panel) of red and blue LSBGs from the new sample. The vertical lines show the median for each class.} 
    \label{fig:color_properties_q}
\end{figure}

Looking at the distribution of \mue{} values of the new sample, both the red and blue LSBGs have a similar mean surface brightness range, with a median \mue{} of 24.75 and  24.68 \magperarcsec{}, respectively. 
Both populations of red LSBGs and blue LSBGs from the new sample have sizes ranging from 2.5\arcsec − 20\arcsec. However, as mentioned earlier, most of these LSBGs have radii less than 7\arcsec{}, with a median of  4.01\arcsec{} for blue LSBGs and 3.59\arcsec{} for red LSBGs. In comparison, blue LSBGs tend to have larger angular radii compared to red LSBGs. 
The S\'ersic index distribution of the red and blue LSBGs in the new sample has similar distribution and almost equal median values (0.847 and 0.845 for red and blue LSBGs, respectively). A median S\'ersic index of around 0.84 indicates that the majority of the galaxies are closer to a disk-shaped geometry, irrespective of their color. The distribution of the axis ratio of the red and blue LSBGs from the new sample shows a clear difference, as shown in Fig.  \ref{fig:color_properties_q}. The median of the axis ratio distribution of the blue and red LSBGs is 0.7 and 0.8, respectively. This indicates that, in general, the red LSBGs are rounder than the blue LSBGs.
 
\subsection{Why are there additional LSBGs?}
\begin{figure}[h]
\centering
  \includegraphics[width=\linewidth]{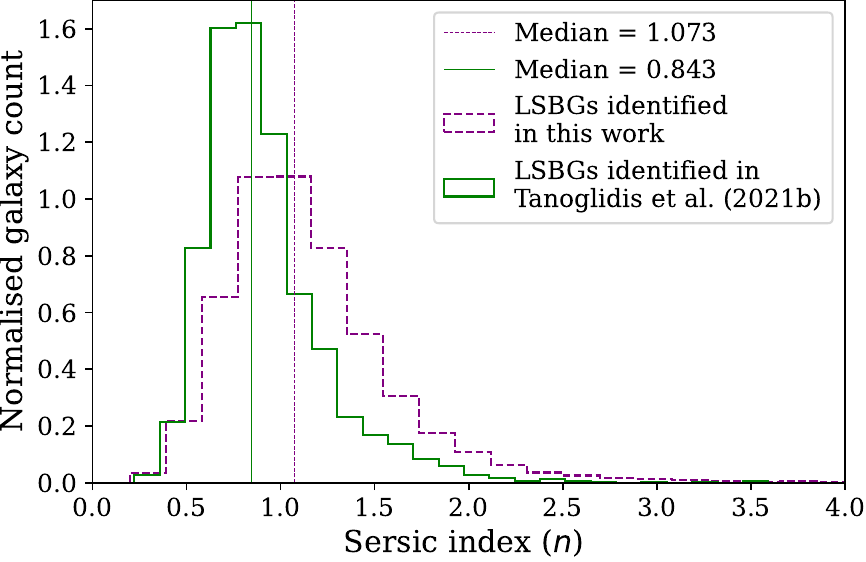}
\caption{Normalised distribution of the S\'ersic index of the LSBGs identified in this work and by \citet{Tanoglidis1}. The vertical lines show the median for each class}.
  \label{fig:compr_n}
\end{figure}

One of the other things to investigate at this moment will be how different the new LSBG sample is compared to the LSBGs identified by \citet{Tanoglidis1}. Or, more precisely, one could wonder why these many LSBGs have been missed previously and whether it is somehow related to the nature of these galaxies. Apart from the S\'ersic index, all other S\'ersic parameters of the new and the old sample have an almost similar distribution. The distribution of the S\'ersic index for the new sample identified in this work and the LSBG sample identified by \citet{Tanoglidis1} is shown in Fig. \ref{fig:compr_n}. The new LSBG sample has a S\'ersic index predominantly in the range $n<1$, which is comparatively lower than the S\'ersic index distribution of LSBGs identified by \citet{Tanoglidis1}. However, this does not point to any reason why these LSBGs were missed in the previous search, and moreover, \citet{Tanoglidis1} have also commented on the under-representation of red LSBGs with small S\'ersic index in their sample. 

\begin{figure*}[h]
\centering
  \includegraphics[width=\linewidth]{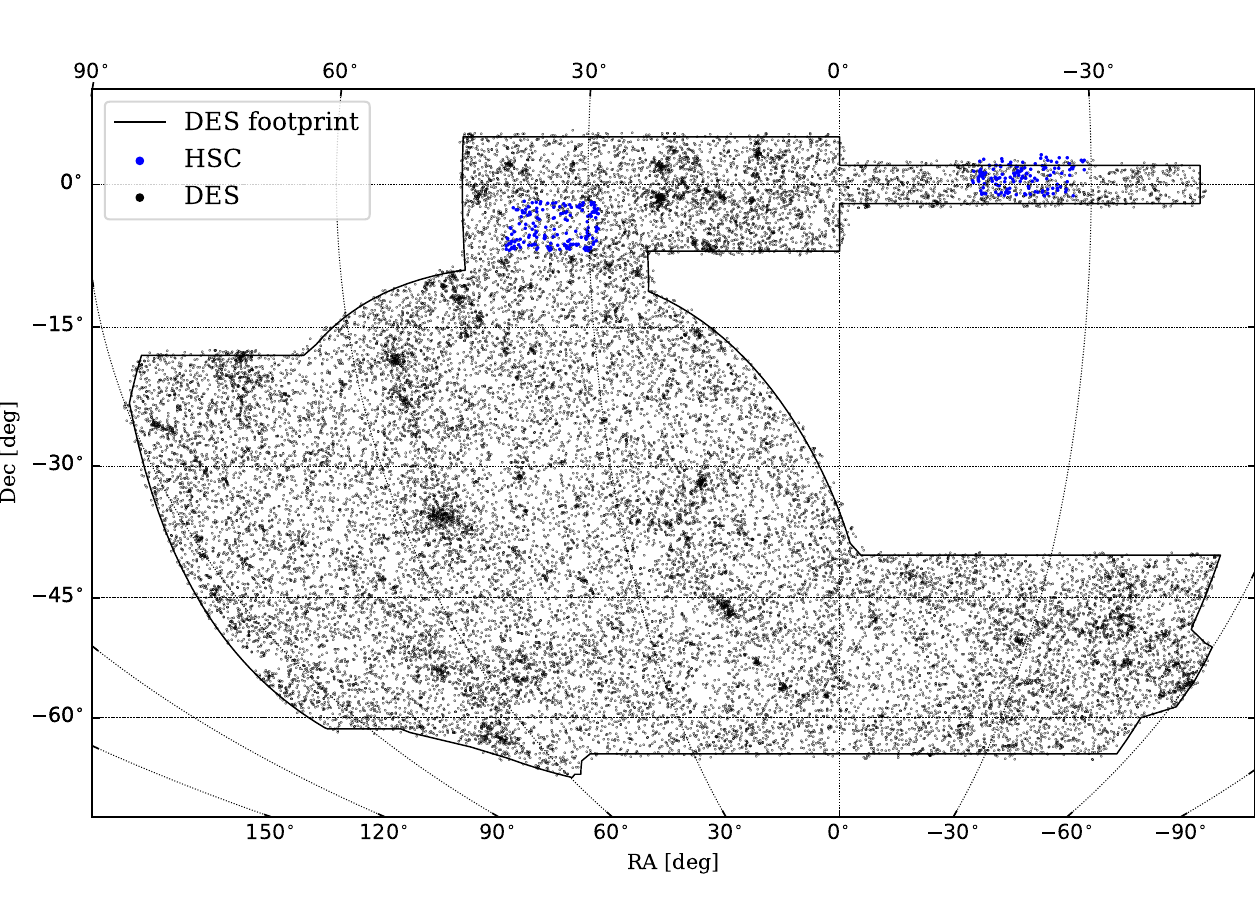}
\caption{Sky distribution of the LSBGs identified from DES (black dots) by \citet{Tanoglidis1} + this work and the LSBGs identified from HSC-SSP (blue dots) by \citet{Greco}.}
    \label{fig:hsc_footprint}
\end{figure*}

To answer the aforementioned question, a close inspection of the methodology of \citet{Tanoglidis1} shows that most of the new LSBGs (82\%) we identified here were missed by the SVM in their first pre-selection step. This shows the importance of methodology in preselecting the samples. Since the methodology used by \citet{Tanoglidis1} and \citet{Greco} have considerable similarities (e.g., usage of SVM), this indicates that \citet{Greco} might have also missed some LSBGs from the HSC-SSP survey and the fraction should be greater in comparison to \citet{Tanoglidis1}. It should be noted that there is a slight overlap in the regions of observation by \citet{Greco} and DES, as shown in Fig. \ref{fig:hsc_footprint}. There are 198 LSBGs identified by \citet{Greco} from HSC-SSP in the field of view of DES and detected in the DES Y3 gold catalogue. Among these 198 LSBGs, \citet{Tanoglidis1} has recovered 183 LSBGs, and we recovered 10 more additional LSBGs from this field, making the total number of recovered LSBGs to 193. We would also like to point out that there are additional LSBGs ($\sim 200$) in our total sample in the same region, but missed by \citet{Greco}, even though the HSC-SSP data used by \citet{Greco} is about 2 orders of magnitude deeper than the DES DR1. However, we have also missed some LSBGs ($\sim 150$) that have been identified by \citet{Greco}. These LSBGs were not detected in the DES Y3 gold catalogue and subsequently were missed by the searches by \citet{Tanoglidis1} and ours. With the DES data release 2 (DES DR 2) having an improved depth ($\sim 0.5$ mag; \citealt{DESDR2}), we should expect an increase in the number of LSBGs from DES. Therefore, there is a potential for using transfer learning with transformers in the future search for LSBGs from DES DR 2 \citep{DESDR2} and HSC-SSP data release 3 \citep{HSCDR3}. 

 With the addition of the new 4\,083 LSBGs, the number of LSBGs in the DES increased to 27,873, effectively increasing the average number density of LSBGs in DES to $\sim 5.5 \text{ deg}^{-2}$. In addition, it should also be noted that there are still around $\sim 3000$ candidates identified by the ensemble models, which have not undergone further analysis to be verified as LSBGs. Potentially indicating that the number of LSBGs in DES might increase further in future. Hence the average number density of $5.5 \text{ deg}^{-2}$ reported here can only be taken as a lower limit. Earlier, \citet{Greco} estimated that the average number density of LSBGs in HSC-SSP is $\sim 3.9 \text{ deg}^{-2}$. However, this estimate was based on LSBG samples with \mue{ $>24.3$} \magperarcsec{}, unlike the \mue{ $>24.2$} \magperarcsec{} selection we adopted in this work. For a similar selection on mue{ $>24.2$} \magperarcsec{} in the combined sample presented here (LSBGs identified in this work + LSBGs identified by \citet{Tanoglidis1}), we obtain a higher number density of 4.9 deg$^{-2}$, compared to the previous estimates (3.9 $\text{ deg}^{-2}$ from \citet{Greco} and 4.5 $\text{ deg}^{-2}$ from \citet{Tanoglidis1}).

\begin{figure}[h]
\centering
  \includegraphics[width=\linewidth]{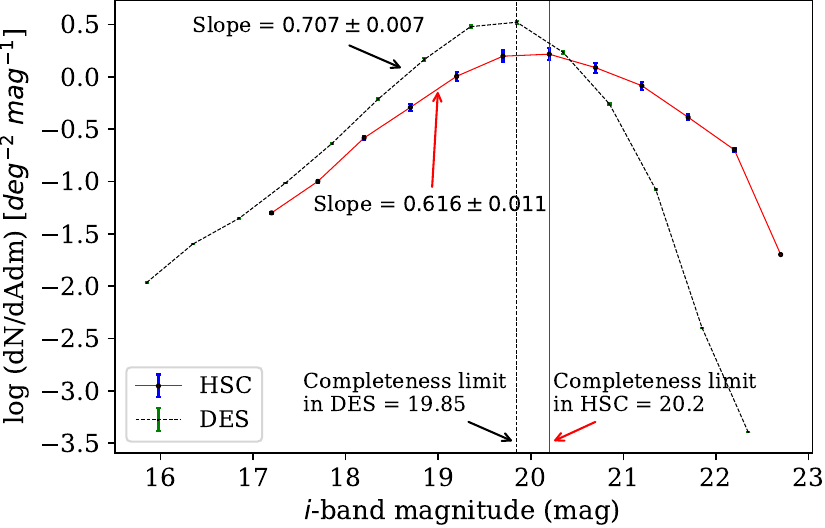}
\caption{The number count of galaxies as a function of $i$-band magnitude, with the y-axis displaying the logarithm of the number density per apparent magnitude. The red line with the blue error bars represents the data from HSC, and the black dashed line with green error bars represents the data from DES.}
    \label{fig:number_count}
\end{figure}

As discussed above, the number density of the LSBGs will be influenced by the methodology used to search for the LSBGs. Similarly, one other intrinsic factor that can influence the number density is the completeness of the survey. Improved imaging techniques can reveal fainter objects, leading to an increase in the number density. The completeness of a survey can be determined by plotting the galaxy number count, and one could also have a rough idea about the redshift distribution of the objects of interest by comparing it with the Euclidean number count. Fig.~\ref{fig:number_count} shows the number count of LSBGs identified in DES (this work and \citet{Tanoglidis1}) and HSC \citep{Greco}. As expected, HSC has better completeness than DES. However, HSC still has a smaller number density than DES, which is evident from comparing the peaks of both number counts. The slope of the number counts near 0.6 (representing Euclidean geometry) for both HSC and DES suggest that most identified LSBGs are local \citep{Yasuda}. Furthermore, \citet{Greene} has analysed the LSBG sample from HSC and estimated that the 781 LSBGs identified by \citet{Greco} have a redshift less than 0.15.

\begin{figure}[h]
\centering
  \includegraphics[width=\linewidth]{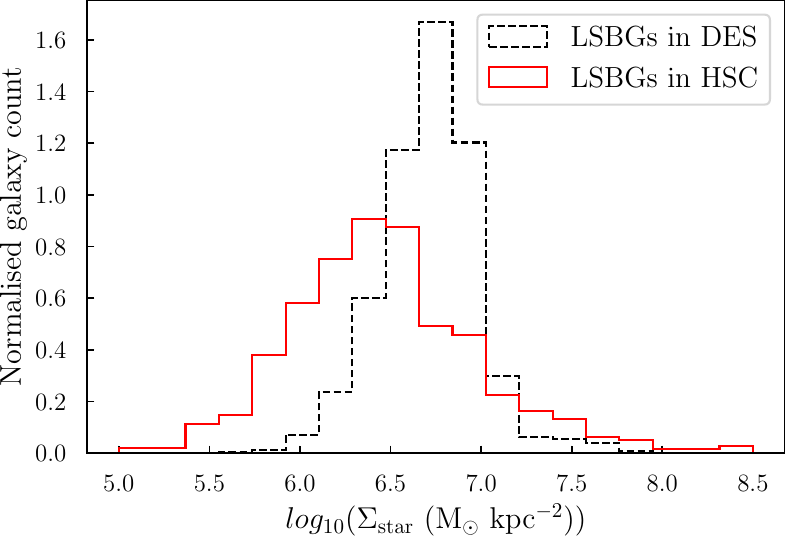}
\caption{Normalised distribution of stellar mass surface density of LSBGs identified in HSC (red line) and DES (black line).}
    \label{fig:s_mass_d_t}
\end{figure}

With the increasing number of LSBGs identified from different surveys, one of the other questions that need to be addressed at this moment is the definition of an LSBG itself. One could define a different definition for an LSBG, consequently leading to finding a completely different sample of LSBGs from the same dataset, which in turn can affect the conclusions of the study. One of the primary factors contributing to these discrepancies is the current reliance on surface brightness-based definitions for LSBGs, which are contingent upon the observation band in use. Different observation bands may involve distinct threshold values. Depending on the band we use, the LSBG definition will likely vary. In this scenario, one potential solution is to define an LSBG based on the stellar mass density of the galaxy. Current definitions based on the stellar mass density define an LSBG as a galaxy with a stellar mass density, $\Sigma_{star} \lesssim 10^7$ M$_{\odot}$ kpc$^{-2}$ \citep[e.g.,][]{Carleton}. Following Eq. 1 of \citet{Chamba2022}, we made an estimate of the stellar mass surface density using our observed \textit{i}-band surface brightness \mue{} and the stellar mass-to-light ratio obtained from the $g-i$ color \citep{Du_2020}. The stellar mass surface density distribution of the LSBGs from DES and HSC-SSP is shown in Fig. \ref{fig:s_mass_d_t}. Here we can see that most of the LSBGs satisfy this condition, and only a small percentage stay above the threshold of $10^7$ M$_{\odot}$ kpc$^{-2}$. On average, the LSBGs from DES have a higher stellar mass surface density than those from HSC-SSP, which could be attributed to the higher depth in the data used by \cite{Greco}. However, as argued by \citet{Chamba2022}, accurate estimation of the stellar mass density requires deep photometry in multiple bands. In our case, we employed a single color, and as a result, the constraints we derived on the stellar mass density may be limited in accuracy.

 \begin{figure*}
\centering
  \includegraphics[width=0.9\linewidth]{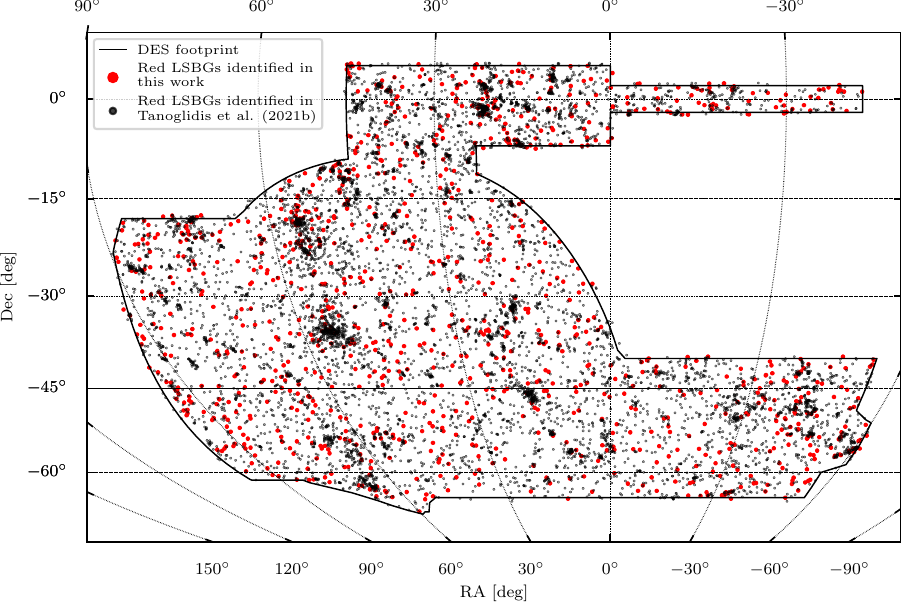}
\caption{Sky distribution of red LSBGs identified in this work (red dots) and the LSBGs identified (black dots) by \citet{Tanoglidis1}.}
\label{fig:red_lbs}
\end{figure*}
\begin{figure*}
\centering
  \includegraphics[width=0.9\linewidth]{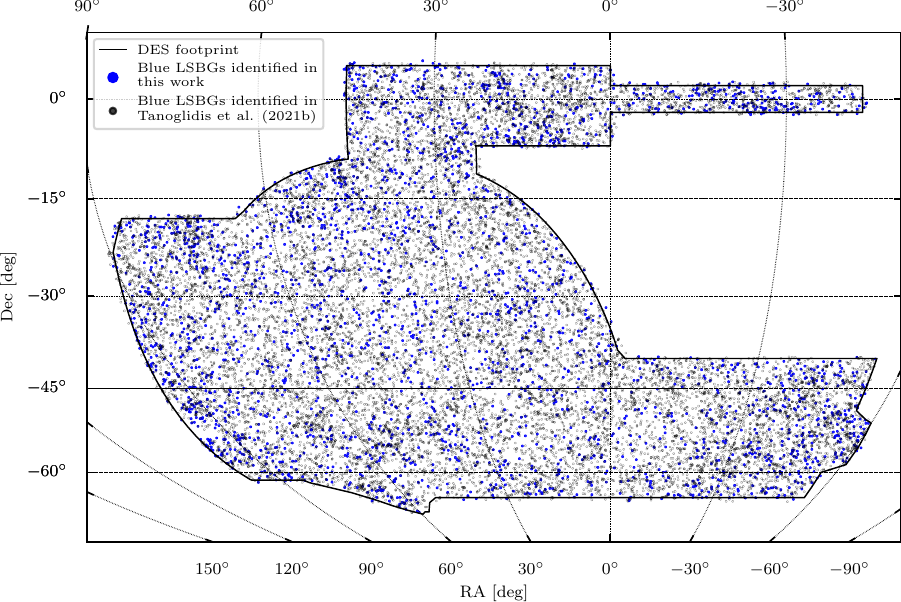}
\caption{Sky distribution of blue LSBGs identified from the new sample (blue dots) and the LSBGs identified (black dots) by \citet{Tanoglidis1}.}
\label{fig:bluelsbs}
\end{figure*}

\section{Clustering of LSBGs in DES}\label{clustering}

The on-sky distribution of the red and blue LSBGs identified in this work, along with those identified by \citet{Tanoglidis1}, is shown in Fig. \ref{fig:red_lbs} and Fig. \ref{fig:bluelsbs}. In the local universe, "normal" high surface brightness red galaxies tend to cluster together, while blue galaxies are much more dispersed in the field \citep{Zehavi2005}. Such a trend is also clearly visible for the LSBG sample. As seen in Fig. \ref{fig:red_lbs}, red LSBGs tend to form concentrated nodes. In contrast, the blue LSBGs are distributed much more homogeneously in the sky, as seen in Fig. \ref{fig:bluelsbs}. 

A two-point auto-correlation function is a statistical tool commonly used to quantify the galaxy clustering \citep{Peebles}. Here we use  the angular two-point auto-correlation function, $\omega (\theta)$, computed using the \citet{Landy} estimator defined as
\begin{equation}
    \omega = \frac{\hat{DD}(\theta)-2\hat{DR}(\theta)+\hat{RR}(\theta)}{\hat{RR}(\theta)},
\end{equation}
where 
\begin{gather}
    \hat{DD} = \frac{DD(\theta)}{n_d(n_d-1)/2},  \\
    \hat{DR} = \frac{DR(\theta)}{n_d n_r},  \\
    \hat{RR} = \frac{DD(\theta)}{n_r(n_r-1)/2}.    
\end{gather}
Here DD($\theta$) is the number of pairs in the real sample with angular separation $\theta$, RR($\theta$)  is the number of pairs within a random sample, DR($\theta$) is the number of cross pairs between the real and random samples, $n_d$ is the total number of real data points, and $n_r$ is the total number of random points.

We use a random sample of \num{4491746} 
points generated from the DES footprint mask. 
To compute $\omega(\theta)$ we employ {\tt treecorr} \citep{Jarvis}. Errors are estimated using jackknife resampling where the sky is divided into 100 equal-sized batches for resampling \citep{Bradley}. For high surface brightness galaxy samples, the angular correlation function very often can be well fitted by a single power-law \citep{Peebles1974, Peebles, Hewett1982, Koo, Neuschaefer}
\begin{equation}
\centering
    \omega (\theta) = A\theta^{1-\gamma} 
\end{equation}
  where $A$ is the amplitude which represents the strength of the clustering, and $\gamma$ represents the rate at which the strength of the clustering reduces as we go to large angular scales. This power-law behaviour is usually observed in a wide range of angular scales; however, it is not universal, especially on the smallest scales. Full modelling of the shape of the correlation function requires taking into account different processes governing galaxy clustering on small scales (corresponding to galaxies located in the same dark matter halo) and at larger scales (corresponding to clustering of different haloes). This modelling is usually done using the halo occupation distribution models (HOD) \citep{Ma, Peacock, Zheng, Kobayashi}. In this work, however, we perform only a preliminary analysis and base interpretation of our data on the power-law fitting only.

  To compare the clustering of the LSBGs with the clustering of the high surface brightness galaxies (HSBGs), we constructed a control sample of HSBGs from the DES data. For this purpose, we selected galaxies in the surface brightness range $20.0 < $ \mue{} $ < 23$ \magperarcsec{} and in the magnitude range $17 < g < 23$ mag (which is the same magnitude range as our LSBG sample). Additionally, we applied a photometric redshift $z < 0.1$ cut in order to keep the HSBGs sample consistent with the LSBGs, which are also expected to be mostly local \citep{Greene}. For this purpose, we used the photometric redshifts from the DES Y3 gold catalogue calculated using the Directional Neighbourhood Fitting (DNF) algorithm \citep{Sevilla, De}. In addition, we also applied the selection cuts on the parameters from {\tt SourceExtractor} such as {\tt  SPREAD\_MODEL},{\tt EXTENDED\_CLASS\_COADD} and on colors (using the  {\tt MAG\_AUTO} magnitudes) as described in Sect. \ref{preselection}.

\begin{figure}[h]
      \centering
      \includegraphics[width=250 pt,keepaspectratio]{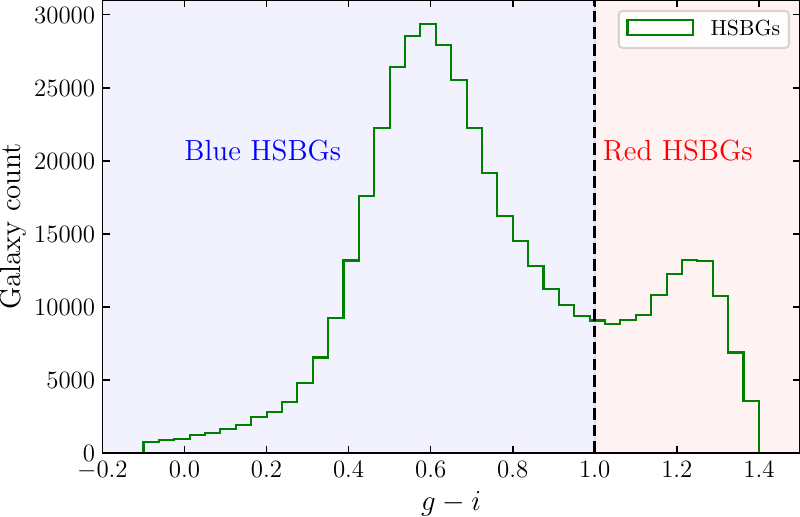}
      \caption{Color distribution of the HSBGs from the DES DR1. The vertical line at $g-i = 1.0$ shows the color separation of the HSBGs into red and blue galaxies.}
      \label{fig:hsbgs_colr}
  \end{figure}

\begin{figure*}[h]
\centering
  \begin{subfigure}{0.475\linewidth}
\centering
\includegraphics[width=\linewidth,keepaspectratio]{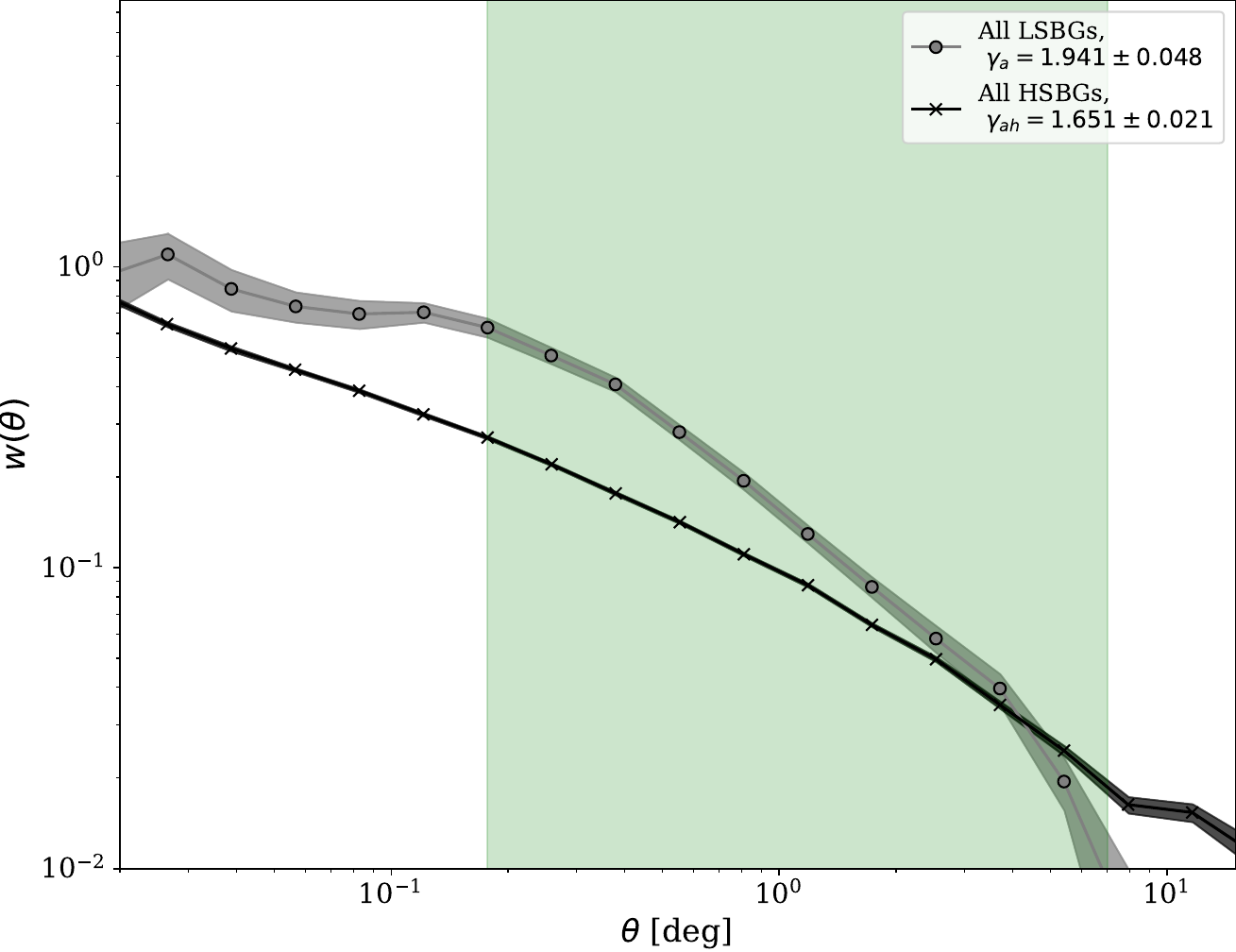}
\label{aaa}
\end{subfigure}
\begin{subfigure}{0.475\linewidth}
\centering
\includegraphics[width=\linewidth,keepaspectratio]{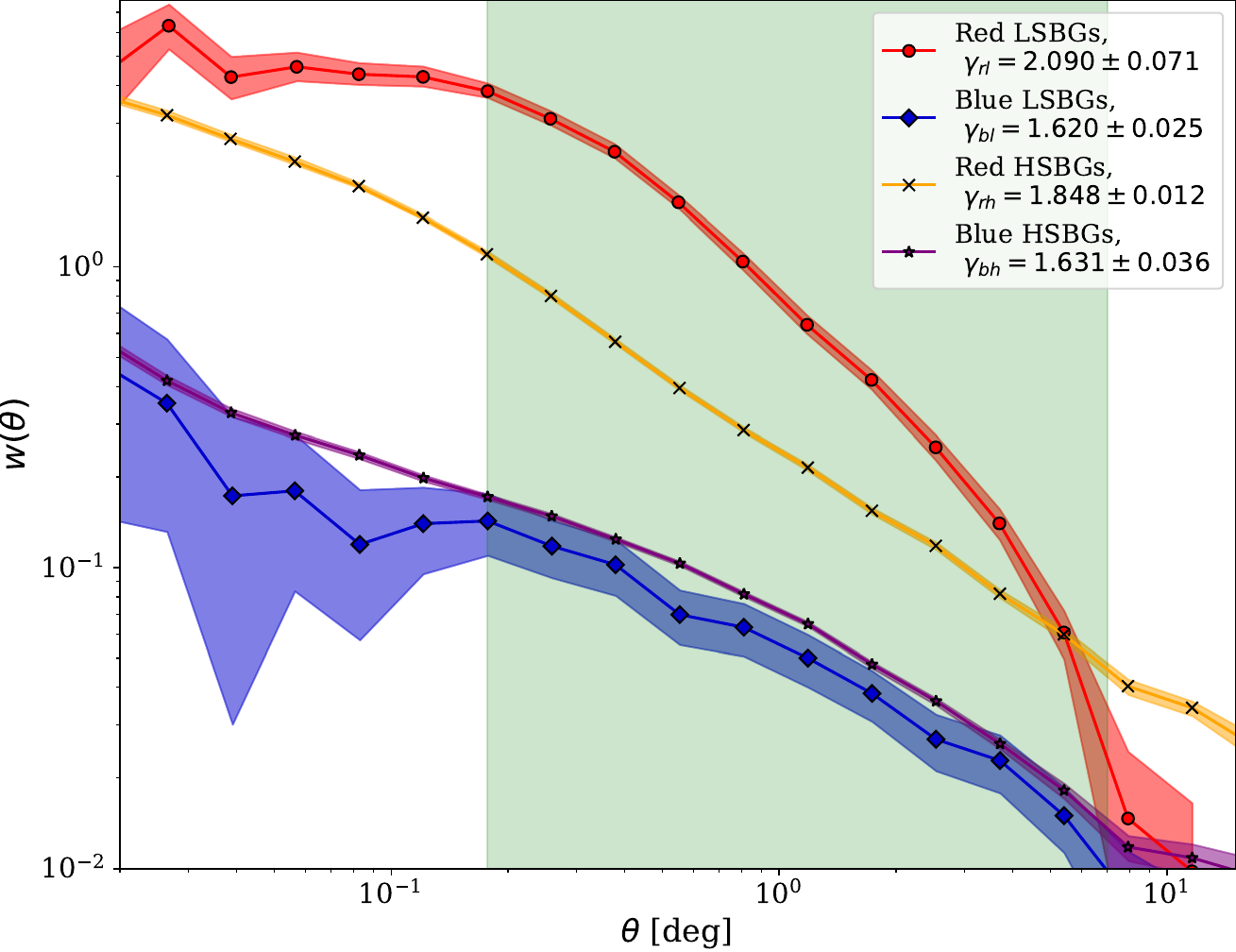}
\label{two}
\end{subfigure}
\caption{Angular autocorrelation function for the full sample of LSBGs (grey line with open circles) and the sample of HSBGs (black line with crosses) is shown in the left panel. The angular autocorrelation function of the red LSBGs (red line), blue LSBGs (blue line), red HSBGs (orange line) and blue HSBGs (purple line) is shown in the right panel. The vertical green shaded region represents the region fitted for a power law ($\omega = A\theta^{1-\gamma}$), and the corresponding $\gamma$ values are shown in the legend. }
    \label{fig:twopointcorrelationfit}
\end{figure*}

  Initially, we computed the angular two-point auto-correlation function for the samples of LSBGs and HSBGs. Then we split the samples into red and blue galaxies to measure their clustering properties separately. For LSBGs, we followed the criterion defined in Sect. \ref{newlsbs}, i.e. a color cut of $g-i=0.6$ mag to separate blue and red sources. As seen from the color histogram presented in Fig.~\ref{fig:hsbgs_colr}, the HSBGs show a bi-modality around $g-i=1.0$ mag, which can be most likely attributed to their different stellar masses. Consequently, we use the boundary  $g-i=1.0$ mag to divide our HSBG sample into red and blue sub-samples.
  The properties of all the samples used for the measurement of the galaxy clustering, together with the best-fit power-law parameters, are listed in Table \ref{corr_fit}. 
  The 2-point auto-correlation functions for all the samples described above are shown in Fig. \ref{fig:twopointcorrelationfit}. 

\begin{table}[ht]
\addtolength{\tabcolsep}{-2.50pt}
\caption{Best-fitting power law parameters for the angular two-point auto-correlation function for HSBG and LSBGs along with the information on the number of galaxies, median $g$-band magnitude, and the mean surface brightness for each sample.}
\begin{tabular}{cccccc}
\toprule
Sample & \makecell{Number of \\ galaxies} & \makecell{Median \\ $g$\\(mag)} & \makecell{Median \\ \mue{}\\(\tiny{\magperarcsec{}})} & \textit{A} & $\gamma$ \\
\midrule
\makecell[l]{All\\HSBGs} & 451,310 & 18.84 & 21.66 & \makecell{0.091\\$\pm$0.004} & \makecell{1.651\\$\pm$0.021} \\
\addlinespace
\makecell[l]{Red\\HSBGs} & 103,900 & 17.96 & 21.21 & \makecell{\textbf{0.245}\\$\pm$\textbf{0.004}} & \makecell{\textbf{1.848}\\$\pm$\textbf{0.012}} \\
\addlinespace
\makecell[l]{Blue\\HSBGs} & 347,410 & 19.21 & 21.81 & \makecell{0.0648\\$\pm$0.004} & \makecell{1.631\\$\pm$0.036} \\
\addlinespace
\hline
\addlinespace
\makecell[l]{All\\LSBGs} & 27,840 & 20.11 & 24.66 & \makecell{0.138\\$\pm$0.013} & \makecell{1.941\\$\pm$0.048} \\
\addlinespace
\makecell[l]{Red\\LSBGs} & 18,924 & 20.23 & 24.89 & \makecell{\textbf{0.671}\\$\pm$\textbf{0.079}} & \makecell{\textbf{2.090}\\$\pm$\textbf{0.071}} \\
\addlinespace
\makecell[l]{Blue\\LSBGs} & 8,916 & 20.07 & 24.59 & \makecell{0.051\\$\pm$0.001} & \makecell{1.620\\$\pm$0.025} \\
\bottomrule
\end{tabular}
\label{corr_fit}
\end{table}

As it is clear from Fig. \ref{fig:twopointcorrelationfit}, the angular two-point auto-correlation function of the red LSBGs does not follow a power law at small angular scales. Therefore, the power-law fits were only performed in the range of $0.15\deg$ to $7 \deg$ to avoid them being affected by the one-halo effects.
In part well fitted by the power law, for the red LSBGs, $\omega(\theta)$ is significantly steeper than for the blue LSBGs.
 However, it flattens at smaller scales, i.e. between $0.01\deg$ and $0.2\deg$; this behaviour is also transmitted to the full sample of LSBGs. In contrast, the blue LSBGs follow a power law behaviour, with a lower clustering amplitude and a much less steep slope, in almost all the angular scales.
 This behaviour of the angular correlation function might be explained by the observations by \citet{Burg} and \citet{Wittmann} that the number of LSBGs close to the cores of galaxy clusters decreases. 
 Such suppression may reduce the clustering power on small scales, leading to a flattening in the auto-correlation function, which is seen for the red LSBGs, which are mostly associated with clusters.

Comparison of clustering of the LSBGs and the HSBGs also shows notable differences. Not surprisingly, red samples, both of HSBGs and LSBGs, are more clustered than their blue counterparts. At the same time, the red LSBG sample has a significantly higher clustering amplitude than the reference red HSBG sample. Red LSBGs also display a steeper slope of $\omega(\theta)$ at angular scales larger than $0.15\deg$, but at smaller scales, their $\omega(\theta)$ flattens, unlike in the case of red HSBGs for which we can even observe a hint of an upturn which can be associated with a one-halo term. This picture is consistent with a scenario in which red LSBGs are mostly associated with dense structures like clusters but do not populate their centres but rather the outskirts. In contrast, red HSBGs display the usual behaviour of red passive galaxies, appearing in a variety of environments, with a tendency to cluster and gather most strongly in the cluster centres.

Blue LSBGs have a significantly lower clustering amplitude than their HSBG counterparts. At the same time, the slope of their $\omega(\theta)$ at scales larger than $0.15\deg$ remains very similar. The blue HSBGs and LSBGs follow the usual distribution of blue star-forming galaxies, dispersed in the field and avoiding clusters. 
These results are consistent with the results obtained by
\citet{Tanoglidis1} for their sample of DES LSBGs. They compared the clustering of LSBGs with very bright galaxies in the magnitude range of $14<g<18.5$ mag from the 2MPZ catalogue \citep{Bilicki}. They found that LSBGs had higher clustering amplitude in the range of 0.1 to 2 degrees, which is similar to our observations. 

However, our results contradict the early estimates from \citet{Bothun_cluster} and \citet{Mo}, who infer that the LSBGs tend to cluster weakly spatially. However, their analyses were limited by a small data sample ($\sim$400 LSBGs), a small area of the sky, and most likely selection biases. 
Given the low accuracy of photometric redshifts for LSBGs in our sample, we do not attempt to reconstruct their spatial clustering in this work. Further analysis is planned as a follow-up to this study.

\section{Identification of ultra-diffuse galaxies}\label{UDGS}
As discussed in Sect. \ref{introduction}, UDGs are a subclass of LSBGs that have extended half-light radii \re{} $ \geq 1.5$ kpc and a central surface brightness $\mu_0 > 24$ \magperarcsec{} in \textit{g}-band \citep{Dokkum}. Significant population of UDGs have been discovered in the Coma cluster by \citet{Dokkum} and other investigations have revealed a large number of UDGs in other galaxy clusters \citep{Koda,Mihos1,Lim,Marca,2022AMarca}. Later on, studies have shown that thousands of UDGs can be found in single individual clusters and that the abundance of UDGs scales close to linearly with host halo mass \citep{Burg, Mancera}.  

In order to investigate if there are any cluster UDGs in the sample of LSBGs we identified in DES, we crossmatched our total LSBG sample ($23,790$ LSBGs from \citealt{Tanoglidis1} and the 4\,083 new LSBGs we identified) with the X-ray-selected galaxy cluster catalogue from the ROSAT All-Sky Survey (RXGCC; \citealt{Xu2022}).  All the LSBGs at the angular distance from the centre of the cluster lower than $R_{200}$\footnote{We used the $R_{500}$ values and the redshifts provided by \citet{Xu2022} to obtain the $R_{200}$ crossmatching radius. Following \citet{Ettori2009}, we assume $R_{200} \approx R_{500}/0.65$ where $R_{500}$ is the radius at which the average density of a galaxy cluster is 500 times the critical density of the universe at that redshift.} virial radius of the cluster were associated with that cluster. Here, $R_{200}$ is the radius at which the average density of a galaxy cluster is 200 times the critical density of the universe at that redshift. We found that 1\,310 LSBGs from the combined catalogue and 123 LSBGs from our new sample were associated with 130 and 53 clusters, respectively. Using the redshift of the cluster provided in \citet{Xu2022}, and assuming that the associated LSBG is at the same redshift as the cluster, we estimated the half-light radius of the LSBG and its projected comoving distance from the cluster centre. It should be noted that since we perform our crossmatching with only projected distances, some of the LSBGs associated with clusters could be non-cluster members that are projected along the field. However, it is unlikely to be the case for all, and since we do not have any other distance estimate for the LSBGs, we chose to adopt this method. However, it should be also noted that UDGs are not exclusively located in clusters; they can also be observed in groups \citep{Cohen,Marleau} and even in field environments \citep{Prole_udgs}. In this section, we are only focusing on the LSBGs and UDGs associated with the clusters.

\begin{figure*}[t]
\centering
\begin{minipage}{0.33\textwidth}
  \includegraphics[width=\linewidth]{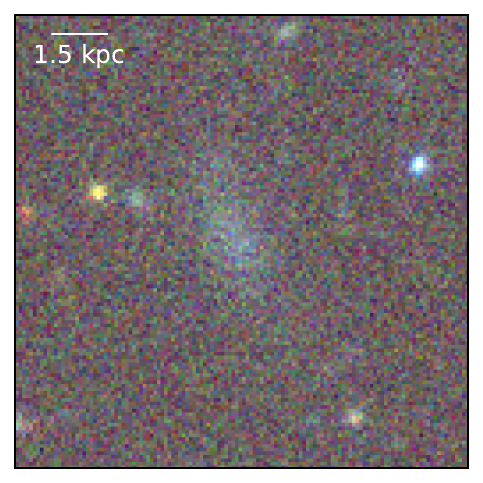}
  \caption*{(a) Coadd Object Id - 221536249}
  \label{fig:udg1}
\end{minipage}\hfill
\begin{minipage}{0.33\textwidth}
  \includegraphics[width=\linewidth]{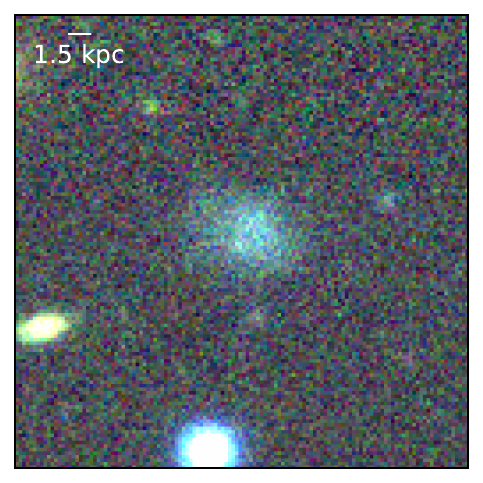}
  \caption*{(b) Coadd Object Id - 287347379}
  \label{fig:udg2}
\end{minipage}\hfill
\begin{minipage}{0.33\textwidth}
  \includegraphics[width=\linewidth]{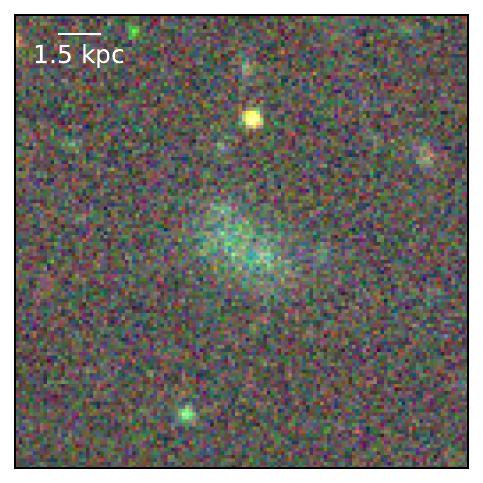}
  \caption*{(c) Coadd Object Id - 323324928}
  \label{fig:udg3}
\end{minipage}

\begin{minipage}{0.33\textwidth}
  \includegraphics[width=\linewidth]{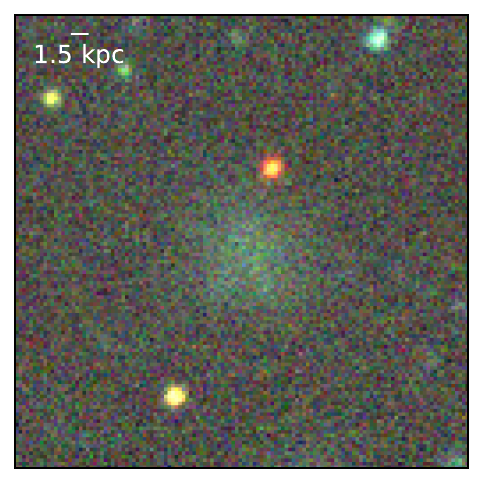}
  \caption*{(d) Coadd Object Id - 392317512}
  \label{fig:udg4}
\end{minipage}\hfill
\begin{minipage}{0.33\textwidth}
  \includegraphics[width=\linewidth]{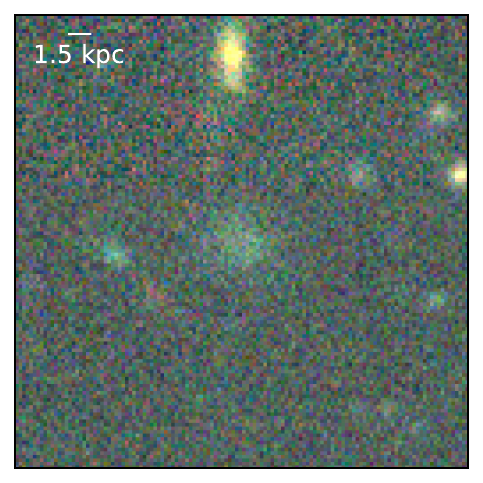}
  \caption*{(e) Coadd Object Id - 295038501}
  \label{fig:udg5}
\end{minipage}\hfill
\begin{minipage}{0.33\textwidth}
  \includegraphics[width=\linewidth]{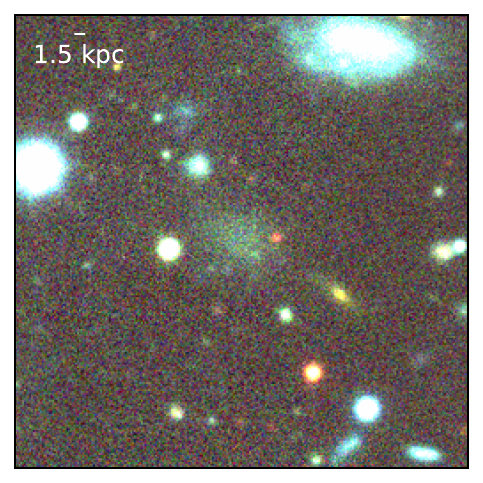}
  \caption*{(f) Coadd Object Id - 461241198}
  \label{fig:udg6}
\end{minipage}

\caption{Cutouts of 6 confirmed new UDGs. The unique identification number (co object id) for each galaxy in DES DR1 is given below each image. The images were generated by combining the $g,r$ and $z$ bands using {\tt APLpy} package \citep{Robitaille}, and each image corresponds to a $33.66 {\arcsec} \times  33.66 {\arcsec}$ region of the sky with the UDG at its centre.}
\label{fig:udgss}
\end{figure*}

Among the 1,310 cluster LSBGs, we further classify 317 cluster UDG candidates based on their half-light radius (\re{}~$\geq1.5$ kpc) and the central surface brightness ($\mu_0$ > 24.0 \magperarcsec{}) in the \textit{g}-band. Since we have not confirmed the physical distances to these galaxies and hence their physical sizes, they can only be regarded as UDG candidates. From here onward, when referring to UDGs in the paper, it is important to note that we are addressing UDG candidates and not confirmed UDGs. These 317  UDGs are distributed within 80 clusters making it the largest sample of clusters in which UDGs are studied.  It should also be noted that \citet{Tanoglidis1} also identified 41 UDGs from their LSBG sample in DES by associating the 9 most overdense regions of LSBGs with known clusters. However, they did not study the properties of those 41 UDGs in detail, and the 276 UDGs among the 317 UDGs reported here are completely new. The UDGs presented here have a median \re{} of 2.75 kpc and $\mu_0$ of 24.51 \magperarcsec{}. Six of the newly identified UDGs are shown in Fig. \ref{fig:udgss}.

\begin{figure}[h]
    \centering
    \includegraphics[width=250 pt,keepaspectratio]{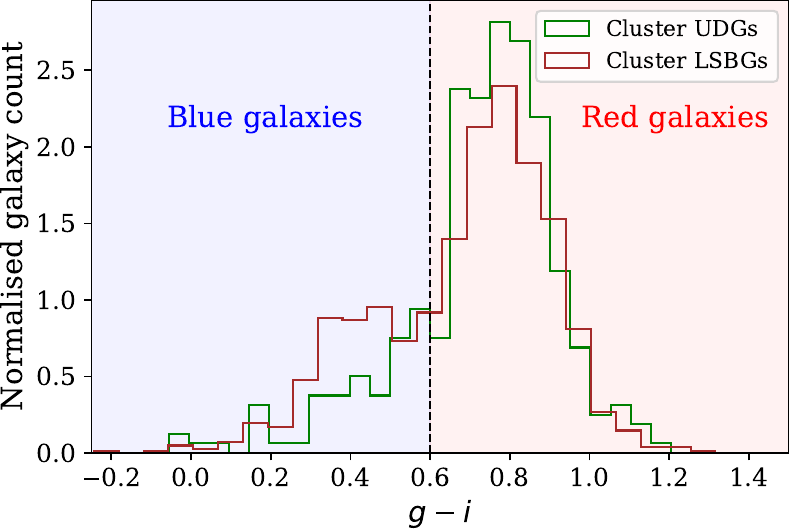}
    \caption{Color distribution of the 1,310 cluster LSBGs and 317 cluster UDGs from the
DES DR1.}
    \label{fig:udg_clr}
\end{figure}

\begin{figure}[h]
    \centering
    \includegraphics[width=250 pt,keepaspectratio]{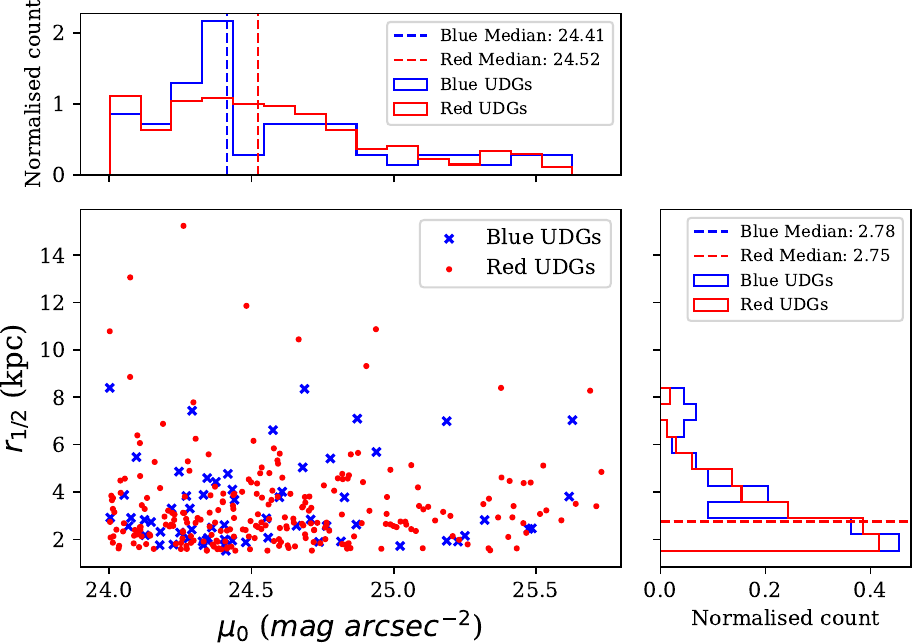}
    \caption{Joint distribution of the red (red dots) and blue (blue cross) UDGs in the space of \re{} and $\mu_0$ in the $g$-band. The vertical lines in the histogram on the x-axis and y-axis show the median for each class. }
    \label{fig:udg_rmue}
\end{figure}

As seen from Fig. \ref{fig:udg_clr}, the majority of the cluster UDGs (253 out of 317) are red in color ($g-i>0.6$ mag), which is similar to the trend of cluster LSBGs (909 out of 1310). This is consistent with theoretical predictions for cluster UDGs \citep{Benavides}. \citet{Mancera_udg} have also found similar distribution for the $g-r$ color of 442 UDGs observed in 8 galaxy clusters. The joint distribution of the red and blues UDGs in the space of \re{} and $\mu_0$ is shown in Fig. \ref{fig:udg_rmue}. The red UDGs presented here have a median \re{} of 2.75 kpc and $\mu_0$ of 24.52 \magperarcsec{}. Similarly,  the blue UDGs have a median \re{} of 2.78 kpc and $\mu_0$ of 24.41 \magperarcsec{}. Most of the red and blue UDGs have a half-light radius in the range 1.5 < \re{} < 6 kpc. However, there is a small fraction of UDGs (6 out of 317) with \re{} > 10 kpc, which is all red and have $\mu_0<25.0$ \magperarcsec{} which might be good potential candidates for the follow-up studies.

\begin{figure*}[h]
\centering
\begin{subfigure}{0.49\linewidth}\label{fig:culster_dist_colr}
\centering
\includegraphics[width=\linewidth,keepaspectratio]{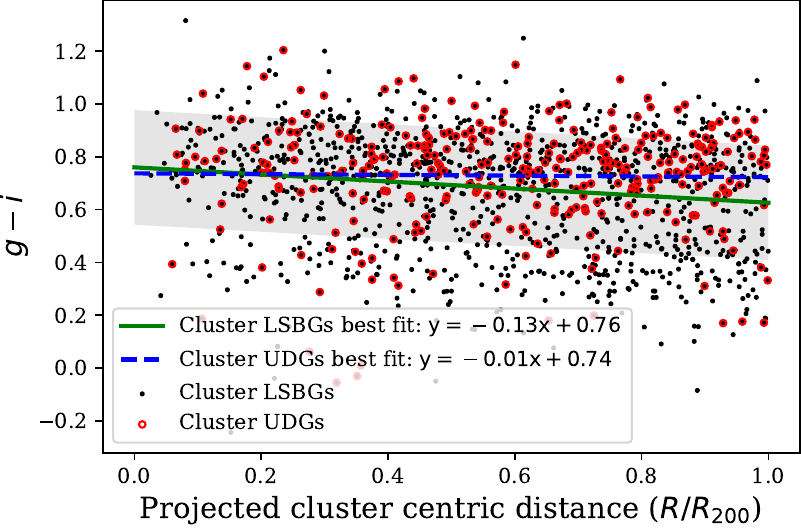}
\end{subfigure}
  \begin{subfigure}{0.49\linewidth}
\centering
\includegraphics[width=\linewidth,keepaspectratio]{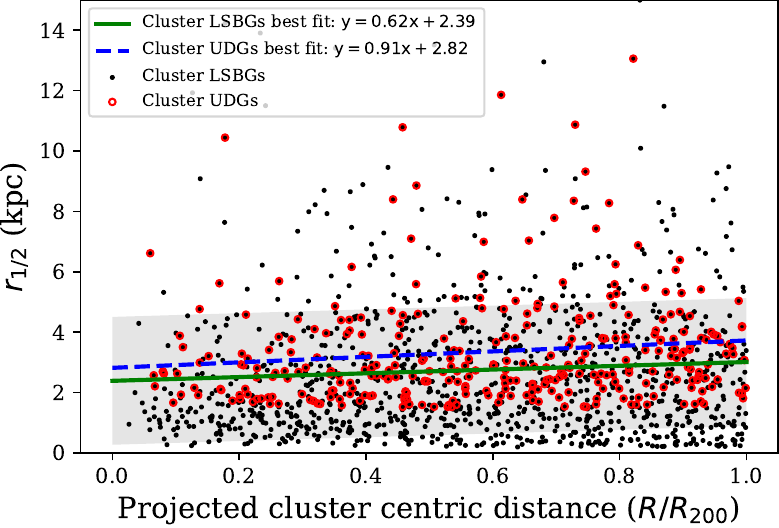}\label{fig:culster_dist_r}
\end{subfigure}
\caption{$g−i$ colour of the cluster LSBGs (black points) and \re{} as a function of the projected distance from their cluster centre (in units of the cluster radius $R_{200}$) is shown in the left and right panel respectively. The UDGs are marked as red hollow circles. The green line and the grey-shaded region are the linear best fit and the $1\sigma$ scatter for the cluster LSBGs, respectively. The blue dashed line is the linear best fit for the cluster UDGs. }
\label{fig:culster_dist}
\end{figure*}

For all the cluster LSBGs, we can see a gradient in color as shown in Fig. \ref{fig:culster_dist}, where LSBGs towards the outskirts of clusters tend to be bluer than those in the centre. This is similar to the behaviour found in Virgo cluster LSBGs from \citet{Junais2022}. However, for the cluster UDGs presented in this study, the color gradient appears much weaker, almost showing a flat distribution in comparison to the LSBGs. A similar weak trend, where more blue UDGs are found towards the cluster centre, was also noted by \citet{Mancera_udg}. On the other hand, \citet{Roman} and \citet{Alabi} reported a more pronounced color trend as a function of cluster-centric distance, while \cite{2022AMarca} did not find any significant trend. However, when directly comparing the trends in the color of UDGs in the cluster, one should keep in mind that these trends will be affected by several factors like the used bands for the color, sample size and the studied cluster, as we can see from the results in the literature, For example, our sample size (>300) is similar to the sample size of \citet{Mancera_udg} and have similar results whereas it is different from the findings of \citet{Roman, Alabi} and \citet{2022AMarca} which have a smaller sample size (<40). 

The trend observed in the half-light radius (Fig. \ref{fig:culster_dist}) for both the cluster LSBGs and UDGs is quite evident. As we move towards the outer regions of the cluster centre, both LSBGs and UDGs show an increase in size. This behaviour is in agreement with the findings of \citet{Roman}. The gradients we observe in color and size with respect to the cluster-centric distance are consistent with the proposed UDG formation scenarios such as the galaxy harassment \citep{Conselice}, tidal interactions \cite{Mancera_udg}, and ram-pressure stripping \citep{Conselice_ram, Buyle}. Such trends are also similar to what is observed for dwarf galaxies in the literature \citep{Venhola}, providing further support for the argument that UDGs can be considered as a subset of dwarf galaxies \citep{Conselice, Benavides}.

The sample of UDG candidates presented here will be the subject of the follow-up analysis. Additionally, it should be noted that all the UDGs reported here are cluster UDGs. The actual number of UDGs in the LSBG catalogue (including low-density environments) might be more than this, and thus the reported number is only a lower limit on the total number of UDGs.

\section{Conclusions}\label{conlclsuin}
In this paper, we explore the possibilities of using transformers in distinguishing LSBGs from artefacts in optical imaging data. 
We implemented four transformer models that combined the use of CNN backbone and self-attention layers to classify the labels;  we call them LSBG DETR (LSBG detection transformers) models. Similarly, we have created four transformer models that directly apply attention to the patches of the images without any convolutions and these models we call LSBG vision transformers. We compared the performance of these two different architectures to the LSBG identification CNN model called DeepShadows presented in \citet{Tanoglidis2}. We found that the transformer models performed better than the DeepShadows, and later we used the ensemble of our transformer models to look for new LSBGs in the DES DR1 data that the previous searches may have missed. 
We follow the definition of an LSBG used by \citet{Tanoglidis1}, i.e. we define LSBGs as galaxies having a \textit{g}-band mean surface brightness $\bar{\mu}_{eff}>24.2$ mag arcsec$^{-2}$ and half-light radii $r_{1/2}>2.5 {\arcsec}$. Following this definition, we identified 4\,083 new LSBGs from the DES DR1, increasing the number of identified LSBGs in DES by 17\%.

Our sample selection and LSBG identification pipeline consist of the following steps: 
\begin{enumerate}
    \item We preselect the objects from the DES Y3 Gold catalog based on the selection criteria described in \citet{Tanoglidis1} using the {\tt SourceExtractor} parameters. 
    \item We applied the ensemble of transformer models to this sample of preselected objects. We chose the objects identified independently by both the LSBG DETR ensemble and the LSBG ViT ensemble for a further follow-up to be inspected for being an LSBG. 
    \item We performed a S\'ersic fitting using  {\tt Galfit} and re-applied the selection cuts to further reduce the number of false positives. 
    After this step, \num{4879} LSBG candidates remained to be visually inspected. 
    \item  After the visual inspection, we report the presence of \num{4083} new LSBGs identified by the transformer ensemble models.
    \end{enumerate}

Following \citet{Tanoglidis1}, we divided the total LSBG sample into two subsamples according to their $g-i$ color. Among the 4083 new LSBGs presented here,  $72\%$ were identified as blue LSBGs, which is higher than the $67\%$ observed in the sample presented by \citet{Tanoglidis1}. Additionally, we also found that we have a more fraction of red LSBGs with color, $g-i > 0.8$, compared to the sample of LSBGs presented by \citet{Tanoglidis1}. We speculate that the bias might have originated from the training set used by \cite{Tanoglidis1} to train the SVM model to preselect the LSBG candidate sample. 

By combining the previously identified 23,790 LSBGs from \citet{Tanoglidis1} with the LSBGs newly identified in our work, the total number of known LSBGs in the DES is increased to 27,873. This increases the number density of LSBGs in the DES from 4.13 to 4.91 deg$^{-2}$ for LSBGs with \mue{} >24.3 \magperarcsec{} and from 4.75 to 5.57 deg$^{-2}$ for LSBGs with \mue{} >24.2 \magperarcsec{}. It should be stressed that this is a lower limit to the number density, and it would increase in the future with better imaging quality and better methodology for the surveys like LSST and Euclid. 

We also made an analysis of the clustering of LSBGs in DES. We found that the LSBGs tend to cluster strongly in comparison to the HSBGs from DES, which is similar to the findings by \citet{Tanoglidis1}. Upon further examination, we observed that the strong clustering tendency observed among low surface brightness galaxies (LSBGs) primarily stems from the red LSBGs, while the behaviour of blue LSBGs resembles that of blue high surface brightness galaxies (HSBGs) with weaker clustering tendencies. Additionally, we noted a decrease in the number of red LSBGs near the centre of the galaxy cluster, resulting in a flattening of the auto-correlation function on smaller scales which is similar to the conclusions of \citet{Wittmann}.

Additionally, we crossmatched the LSBGs with the X-ray-selected galaxy cluster catalogue from the ROSAT All-Sky Survey (RXGCC; \citealt{Xu2022}) to find LSBGs associated with the clusters. Using the redshift information of the clusters, we identify 317 UDGs, among which 276 are reported for the first time. We also observed a color gradient among the cluster LSBGs, where LSBGs located towards the outskirts of clusters exhibit a bluer color compared to those at the centre, similar to findings by \citet{Junais2022} in the Virgo cluster LSBGs. However, this trend is relatively weak for the cluster UDGs in our study, unlike the LSBGs. A clear trend in the half-light radius of the cluster LSBGs and UDGs as a function of the cluster-centric distance is also visible. The LSBGs and UDGs grow in size as going from the cluster centre to the outskirts. The coherent trends in the color and size are in agreement with the proposed UDG formation mechanisms such as the galaxy harassment \citep{Conselice}, tidal interactions \cite{Mancera_udg}, and ram-pressure stripping \citep{Conselice_ram, Buyle}, giving more support to the argument that the UDGs are a subset of dwarf galaxies \citep{Conselice, Benavides}.

The upcoming large-scale surveys such as LSST and Euclid are expected to cover around 18\,000 and 14\,5000 deg$^2$ of the sky, respectively \citep{Ivezi__2019,scaramella2021euclid}. Extrapolating our results on the number density of LSBGs, we are expected to find more than 100\,000 and 80\,000 LSBGs from LSST and Euclid, respectively. In this scenario, an improved and efficient methodology will be highly significant, and we propose that transformer models could overcome this difficulty. With the aid of transfer learning, we are planning to extend our study to HSC SSP DR3 and thus pave a pathway for the LSBG detection in LSST and Euclid.

\begin{acknowledgements}
J and KM are grateful for support from the Polish National Science Centre via grant UMO-2018/30/E/ST9/00082. US acknowledges support from the National Research Foundation of South Africa (grant no. 137975). This research was partially supported by the Polish National Science Centre grant UMO-2018/30/M/ST9/00757 and the Polish Ministry of Science and Higher Education grant DIR/WK/2018/12. 
\end{acknowledgements} 

\bibliographystyle{aa}
\bibliography{aa}

\end{document}